\documentclass[aps,prb,floatfix,twocolumn,superscriptaddress]{revtex4-2}
\usepackage[bookmarks=false,colorlinks=true,linkcolor=blue,filecolor=blue,citecolor=blue,urlcolor=blue]{hyperref}
\usepackage{amsmath,amssymb,amsthm}
\usepackage{graphicx} 
\usepackage{bm}
\usepackage{epsfig}
\usepackage{times}
\usepackage{color}
\usepackage[pdftex,dvipsnames]{xcolor}
\usepackage[normalem]{ulem}
\graphicspath{{./figures/}{./}}

\begin{document}

\title{Phase transitions in coupled Ising chains and SO($N$)-symmetric spin chains}

\author{Yohei Fuji}
\affiliation{Department of Applied Physics, University of Tokyo, Tokyo 113-8656, Japan}
\affiliation{Department of Physical Sciences, Ritsumeikan University, Kusatsu, Shiga 525-8577, Japan}

\author{Sylvain Capponi}
\affiliation{Univ Toulouse, CNRS, Laboratoire de Physique Th\'eorique, Toulouse,  France.}

\author{Lukas Devos}
\affiliation{Center for Computational Quantum Physics, Flatiron Institute, New York 10010, USA}

\author{Philippe Lecheminant}
\affiliation{Laboratoire de Physique Th\'{e}orique et Mod\'{e}lisation, CNRS, CY Cergy Paris Universit\'{e}, 95302 Cergy-Pontoise Cedex, France}

\date{\today}

\begin{abstract}
We investigate the nature of quantum phase transitions in a (1+1)-dimensional field theory composed of $N$ copies of the Ising conformal field theory interacting via competing relevant perturbations.
The field theory governs the competition between a mass term and an interaction involving the product of $N$ order-parameter fields, which is realized, e.g. in coupled Ising chains, two-leg spin ladders, and SO($N$)-symmetric spin chains.
By combining a perturbative renormalization group analysis and large-scale matrix-product state simulations, we systematically determine the nature of the phase transition as a function of $N$.
For $N=2$ and $N=3$, we confirm that the transition is continuous, belonging to the Ising and four-state Potts universality classes, respectively.
In contrast, for $N \ge 4$, our results provide compelling evidence that the transition becomes first order.
We further apply these findings to specific lattice models with SO($N$) symmetry, including spin-$1/2$ and spin-$1$ two-leg ladders, that realize a direct transition between an SO($N$) symmetry-protected topological phase and a trivial phase.
Our results refine a recent conjecture regarding the criticality of transitions between SPT phases.
\end{abstract}

\maketitle
\tableofcontents

\section{Introduction}

One-dimensional (1D) quantum systems have been an ideal platform to investigate physical phenomena driven by strong quantum fluctuations \cite{Giamarchi, Gogolin}.
These fluctuations often lead to the emergence of novel phases and critical phenomena, whose universal behavior is elegantly captured by (1+1)-dimensional conformal field theory (CFT) \cite{CFT}.
As a powerful nonperturbative tool, CFT allows us to classify a diverse range of 1D criticality and to exactly compute universal quantities, such as the central charge and critical exponents, which govern the asymptotic behaviors of various physical observables.
Furthermore, by treating CFT as a critical fixed point, one can employ the renormalization group (RG) to study the effect of relevant perturbations and to determine the nature of phases near the critical point \cite{Zamolodchikov-c-87,Cardy, Mussardo}.

While it is often straightforward to analyze the RG flow of a single relevant operator, typically leading to a massive phase, the situation becomes far more intricate when two or more relevant operators are present.
In such cases, the competition between these operators can stabilize a nontrivial critical fixed point, which describes a continuous phase transition between the distinct phases favored by each operator.
Well-known examples of this mechanism include the emergence of the Wilson-Fisher fixed point in the $\phi^4$ theory \cite{Wilson72, Wilson74} and the bicritical point associated with spin-flop transitions in anisotropic antiferromagnets \cite{Nelson74, Kosterlitz76}. 
There are several examples of (1+1)-dimensional field theories that correspond to a CFT perturbed by two
strongly relevant operators with antagonistic effects.
Each relevant perturbation gives a massive quantum field theory with different ground states, but the interplay between the two competing relevant operators may produce a quantum phase transition that is described by a CFT.
Although there are specific cases, such as the two-frequency sine-Gordon model \cite{Delfino98,Fabrizio00,Bajnok01} and self-dual sine-Gordon models \cite{Lecheminant02, Delfino2002, Lecheminant06, Lecheminant07, Fuji-L-17}, where the infrared CFT is known, and Zamolodchikov's $c$-theorem restrict the possible infrared theories by ensuring that the central charge must decrease along the RG flow \cite{Zamolodchikov86}, a complete characterization of the emergent CFT generally remains a challenging task.

In this manuscript, we investigate such a nontrivial RG flow by considering a (1+1)-dimensional field theory described by $N$ copies of the Ising CFT, 
\begin{align} \label{eq:NIsingCFTHamintro}
H &= -\frac{iv}{4\pi} \int dx \sum_{a=1}^N (\xi_R^a \partial_x \xi_R^a -\xi_L^a \partial_x \xi^a_L) \nonumber \\
&\quad -im\int dx \sum_{a=1}^N \xi_R^a \xi_L^a +\lambda_1 \int dx \prod_{a=1}^N \sigma_a ,
\end{align}
where $\xi^a_R$ ($\xi^a_L$) are right-moving (left-moving) Majorana fermion fields and $\sigma_a$ are Ising spin fields corresponding to order parameters.
It describes $N$ copies of Ising CFTs interacting via two perturbations $m$ and $\lambda_1$.
The vanishing mass $m=0$ and vanishing $N$-spin interaction $\lambda_1=0$ give a critical point described by $N$ copies of the Ising CFTs with central charge $c=N/2$.
The $m$ term represents the thermal operator for each Ising CFT and is a relevant perturbation with scaling dimension 1.
For a finite mass, the system with  $\lambda_1=0$ lies in an ordered phase with $\langle \sigma_a \rangle \neq 0$ for $m<0$ and in a disordered phase with $\langle \sigma_a \rangle =0$ for $m>0$.
The $\lambda_1$ term is the product of $N$ Ising spin fields and has a scaling dimension $N/8$, which is relevant for $N<16$. The field theory in Eq.~\eqref{eq:NIsingCFTHamintro} describes the competition between two strongly relevant operators that favor different phases: the mass term that stabilizes a disordered phase for $m>0$ and the $N$-spin interaction that promotes an ordered phase.
Such interplay may produce a quantum phase transition, possibly described by a non-trivial CFT, between the two expected phases.

The nature of the quantum phase transition in the field theory given by Eq.~\eqref{eq:NIsingCFTHamintro} has already been studied in the $N=2$ and $N=3$ cases, where the theory describes the competition between different orders in spin chains, 1D interacting electrons, and ladder models.
For instance, the theory in Eq.~\eqref{eq:NIsingCFTHamintro} with $N=2$ governs the transition from a band insulator to a dimerized phase in the 1D Hubbard model with an alternating potential, as well as the transition from a N\'{e}el ordered phase to a dimerized phase in a spin-$1/2$ Heisenberg chain with a staggered magnetic field \cite{Fabrizio99, Fabrizio00}.
In this case, an Ising critical point with central charge $c=1/2$ has been found, which stems from the interplay between the two strongly relevant operators in Eq.~\eqref{eq:NIsingCFTHamintro} \cite{Fabrizio00}.
Similarly, the theory with $N=3$ governs the quantum phase transition in a two-leg spin-$1/2$ Heisenberg ladder or a spin-$1$ Heisenberg chain both with explicit dimerization, for which an emergent SU(2)$_1$ quantum criticality with $c=1$ has been predicted \cite{YJWang00}.

Building upon these prior results, we show that the effective field theory in Eq.~\eqref{eq:NIsingCFTHamintro} is rather generic in one dimension; it governs the quantum phase transitions between different ordered phases in spin ladder models, as well as those between a nondegenerate featureless phase and an SO($N$) symmetry protected topological (SPT) phase \cite{Duivenvoorden-Q-13, Tu2008}.
In the latter case, it has been conjectured that the critical point between a SPT phase with $d$-fold degenerate edge state and a trivial phase is described by a CFT with central charge $c \ge \log_2 (d)$, which stems from the delocalization of the edge state at the transition \cite{Verresen17}.
We here systematically study the nature of phase transitions in the field theory given by Eq.~\eqref{eq:NIsingCFTHamintro} as a function of $N$ by combining a perturbative RG analysis and large-scale numerical simulations using matrix-product states (MPS) for different lattice models in the context of spin ladders, SO($N$) spin chains, and coupled Ising chains.
We provide compelling evidence that the transition is continuous for $N=2$ (Ising universality) and $N=3$ (four-state Potts universality), in precise agreement with the predictions of Refs.~\cite{Fabrizio00, YJWang00} based on a strong-coupling expansion, but becomes first-order for all $N \ge 4$.
This suggests the existence of a critical value $3 < N_c < 4$, above which a continuous transition is no longer possible. 
In particular, we find that the phase transition between an SO($N$) SPT phase and a trivial phase is generically first-order  when $N$ is odd and $N \ge 5$ and is not described by a CFT, in contrast to the assumption behind the conjecture in Ref.~\cite{Verresen17}.

The rest of the manuscript is structured as follows.
In Sec.~\ref{sec:CoupledIsingCFTs}, we present the CFT description of the theory in Eq.~\eqref{eq:NIsingCFTHamintro} as well as a one-loop RG approach using an $\epsilon = 16 -N$ expansion.
A lattice realization of the field theory in Eq.~\eqref{eq:NIsingCFTHamintro} in terms of $N$ coupled Ising chains is given in Sec.~\ref{sec:CoupledIsing}, which enables us to investigate the phase transition in the $N=2,3,4$ cases by means of MPS calculations.
Section~\ref{sec:SPTPhaseTransitions} focuses on phase transitions in spin ladders and SO($N$) spin chains, which are described by the field theory in Eq.~\eqref{eq:NIsingCFTHamintro} for $N=2$ to $N=6$. 
Large-scale MPS calculations are also used to investigate the nature of the phase transition in these models. 
A summary of the main results is given in Sec.~\ref{sec:Conclusion}, and the manuscript is supplied with two appendices to provide technical details and additional information.

\section{Coupled Ising CFTs}
\label{sec:CoupledIsingCFTs}

We consider the following Hamiltonian defined in one spatial dimension, which describes $N$ coupled Ising CFTs,
\begin{align} \label{eq:NIsingCFTHam}
H &= -\frac{iv}{4\pi} \int dx \sum_{a=1}^N (\xi_R^a \partial_x \xi_R^a -\xi_L^a \partial_x \xi^a_L) \nonumber \\
&\quad -im\int dx \sum_{a=1}^N \xi_R^a \xi_L^a +\lambda_1 \int dx \prod_{a=1}^N \sigma_a \nonumber \\
&\quad -\lambda_2 \int dx \sum_{a<b} \xi_R^a \xi_L^a \xi_R^b \xi_L^b.
\end{align}
This is the field theory in Eq.~\eqref{eq:NIsingCFTHamintro} with an additional marginal interaction with coupling constant
$\lambda_2$ among the Majorana fermion fields, which will be generated in the RG approach.
It describes the $N$ copies of Ising CFTs interacting via three perturbations $m$, $\lambda_1$, and $\lambda_2$ with scaling dimensions $\Delta_m = 1$, $\Delta_1 = N/8$, and $\Delta_2=2$, respectively.

For $\lambda_1=\lambda_2=0$, the Hamiltonian in Eq.~\eqref{eq:NIsingCFTHam} describes $N$ decoupled Majorana fermions with identical mass $m$. 
The critical point coincides with the vanishing mass $m=0$ and corresponds to $N$ copies of the Ising CFTs with central charge $c=N/2$.
For a finite mass, the system is in an ordered (disordered) phase with $\langle \sigma_a \rangle \neq 0$ ($\langle \sigma_a \rangle =0$) when $m<0$ ($m>0$).

Let us suppose $N<16$ so that a finite $\lambda_1$ gives another relevant perturbation on top of the mass term.
The $\lambda_1$ term prefers to acquire finite expectation values for the order parameter fields $\sigma_a$.
In the ordered phase with $m<0$, this is compatible with $\langle \sigma_a \rangle \neq 0$ and does not have any dramatic effect.
However, in the disordered phase $m>0$, the $\lambda_1$ term competes with the mass term and a strong $\lambda_1$ will lead to a phase transition to an ordered phase with $\langle \sigma_a \rangle \neq 0$. 
In the $N=2,3$ cases, Refs.~\cite{Fabrizio00,YJWang00} suggested that the transition is continuous by means of a strong-coupling expansion.
A natural question is whether this non-trivial infrared (IR) criticality persists for $N \ge 4$.

\subsection{RG analysis} \label{sec:RG}

A first possible approach is to derive the perturbative RG equations at one loop for the field theory in Eq.~\eqref{eq:NIsingCFTHam}
near the non-interacting point with $c=N/2$.
The approach is obviously not fully controlled since the scaling dimensions of some perturbations are not small. 
However, it might shed light on the possible IR behavior of the theory by considering the vicinity of $N=16$, where the $\lambda_1$ perturbation in Eq. \eqref{eq:NIsingCFTHam} becomes marginal.
Using the operator product expansion for the Ising CFT, we can write down perturbative RG equations for the coupling constants $G_0 \propto m$, $G_1 \propto \lambda_1$, and $G_2 \propto \lambda_2$. 
As detailed in Appendix~\ref{app:DerivationRGEq}, we obtain the one-loop RG equations 
\begin{subequations} \label{eq:RGEq}
\begin{align}
\frac{dG_0}{dl} &= G_0 -\frac{1}{4} G_1^2 -(N-1) G_0 G_2 \\
\frac{dG_1}{dl} &= \left( 2-\frac{N}{8} \right) G_1 -\frac{N}{2} G_0 G_1 -\frac{N(N-1)}{8} G_1 G_2, \\
\frac{dG_2}{dl} &= -G_0^2 -\frac{1}{8} G_1^2 -(N-2) G_2^2.
\end{align}
\end{subequations}

To study fixed-point properties of these RG equations, let us first consider the RG equations only for $G_0$ and $G_1$ and neglect the marginal coupling $G_2$ for a moment.
Since $G_1$ is marginal at $N=16$, we could employ $\epsilon$-expansion by setting $N=16-\epsilon$ to perturbatively analyze nontrivial fixed points. 
Indeed, the RG equations for $G_0$ and $G_1$ become
\begin{subequations}
\begin{align}
\frac{dG_0}{dl} &= G_0 -\frac{1}{4} G_1^2, \\
\frac{dG_1}{dl} &= \frac{\epsilon}{8} G_1 -\left( 8-\frac{\epsilon}{2} \right) G_0 G_1,
\end{align}
\end{subequations}
and they have two nontrivial fixed points, 
\begin{align} \label{eq:FixedPoints}
(G_0^*, G_1^*) = \left( \frac{\epsilon}{64} +O(\epsilon^2), \pm \frac{\sqrt{\epsilon}}{4} +O(\epsilon^{3/2}) \right).
\end{align}
These fixed points will correspond to the transition between a disordered phase for $0 < |\lambda_1| \ll m$ and an ordered phase for $0 < m \ll |\lambda_1|$. 

However, these fixed points do not necessarily imply the existence of a continuous (scale-invariant) phase transition, since a finite $G_2 < 0$ is inevitably generated under the RG transformation even if we initially set $G_2(l=0)=0$.
Since $G_2$ is marginally relevant, we need to consider the full RG equations in Eq.~\eqref{eq:RGEq} to investigate the true nature of the phase transition. 
It turns out that $\epsilon$-expansion of the full RG equations only yields nontrivial fixed points in the complex parameter space of $G_j$'s, as given in Appendix~\ref{app:FixedPointRGEq}.
Thus, there are no scale-invariant fixed points perturbatively accessible for small $\epsilon = 16-N$, meaning that the transition for $N$ slightly below $16$ will only be a first-order transition. 

These arguments, of course, do not forbid the existence of a continuous phase transition for $N \ll 16$, where the perturbative analysis breaks down. 
Indeed, a continuous phase transition has been found for $N=2$ \cite{Fabrizio00} and $N=3$ \cite{YJWang00}.
This indicates that there is a threshold value of $N=N_c \in (3,16)$, above which the phase transition becomes the first-order type. 
While an early study \cite{YJWang03} suggested that the $N=4$ model also has a continuous phase transition, we argue below that the threshold value is in the range $3 < N_c < 4$ and numerically show that the theory in Eq.~\eqref{eq:NIsingCFTHam} for $N \geq 4$ always has a first-order transition.

\subsection{SO($N$)$_1$ CFT description}

The Hamiltonian in Eq.~\eqref{eq:NIsingCFTHam} has an SO($N$) symmetry and can equivalently be described by the SO($N$)$_1$ Wess-Zumino-Witten (WZW) CFT \cite{CFT}. 
This CFT is generated by the right (left) SO($N$)$_1$ currents $J^{ab}_R$ ($J^{ab}_L$) with $1 \leq a < b \leq N$, which can be expressed in terms of the Majorana fermion fields as $J^{ab}_R = i\xi^a_R \xi^b_R$ ($J^{ab}_L = i\xi^a_L \xi^b_L$). 
The SO($N$)$_1$ currents transform in the adjoint representation of SO($N$) so that the combinations like $:\mathrel{\bm{J}_R \cdot \bm{J}_R}:$ or $\bm{J}_R \cdot \bm{J}_L$ are SO($N$) singlets, where we have used the vector notation $\bm{J}_{R/L} = (J^{ab}_{R/L})_{1 \leq a < b \leq N}$ and $:\mathrel{X}:$ is the normal-ordered product of $X$.
The Majorana fermion fields $\xi^a_R$ and $\xi^a_L$ correspond to chiral primary fields of conformal weight $1/2$ and transform in the vector representation of SO($N$).
They are combined to form an SO($N$) singlet $\sum_a \xi^a_R \xi^a_L$, which is nothing but the mass term in Eq.~\eqref{eq:NIsingCFTHam}.
Since the marginal term proportional to $\lambda_2$ can be written as a product of these singlets or as $\bm{J}_R \cdot \bm{J}_L$ using the SO($N$)$_1$ currents, it is an SO($N$) singlet.
The $\lambda_1$ term in Eq.~\eqref{eq:NIsingCFTHam} can be expressed in terms of the SO($N$)$_1$ WZW primary field $\Phi_\mathrm{S}$ with scaling dimension $N/8$, which transforms in the spinor representation of SO($N$). 
We note that $\Phi_\mathrm{S}$ is unique in the odd-$N$ case, whereas it is a suitable combination of the primary fields associated with two irreducible representations of the spinor representation in the even-$N$ case. 
In terms of these fields, the Hamiltonian in Eq.~\eqref{eq:NIsingCFTHam} can be rewritten as
\begin{align}
H &= \frac{v}{4\pi (N-1)} \int dx \, (:\mathrel{\bm{J}_R \cdot \bm{J}_R}:+:\mathrel{\bm{J}_L \cdot \bm{J}_L}:) \nonumber \\
&\quad -im\int dx \sum_{a=1}^N \xi_R^a \xi_L^a +\lambda_1 \int dx \, \mathrm{Tr} (\Phi_{\rm S}+\Phi^\dagger_\mathrm{S})  \nonumber \\
&\quad +\lambda_2 \int dx \, \bm{J}_R \cdot \bm{J}_L, 
\end{align}
which gives an explicit SO($N$) invariant formulation of the theory in Eq.~\eqref{eq:NIsingCFTHam}.

\subsection{Correspondence with lattice models}

Both descriptions in terms of $N$ copies of Ising CFT and SO($N$)$_1$ WZW CFT are equivalent at the level of field theory. The nature of the phase transition, continuous or first-order, is not affected by the choice of these descriptions.
However, once we consider a specific lattice model to realize the field theory in Eq.~\eqref{eq:NIsingCFTHam}, we need to care about the correspondence between local operators on the lattice (such as spin operators on each lattice site) and primary fields in order to correctly interpret the resulting phases and phase transition.

We first consider the case where there exist local lattice operators corresponding to each of the Ising order-parameter fields $\sigma_a$, as well as thermal (energy) fields $\xi^a_R \xi^a_L$.
In other words, the operator content of the field theory in Eq.~\eqref{eq:NIsingCFTHam} matches that of $N$ copies of the Ising CFT in this case.
Here, the phase driven by $\lambda_1$ corresponds to an ordered phase with spontaneous breaking of the $\mathbb{Z}_2^{N-1}$ symmetry whereas that by positive $m$ to a symmetric disordered phase. 
We will study the associated lattice model in depth in the next section and argue that the continuous phase transition for $N=2$ and $N=3$ belongs to the Ising and four-state Potts universality class, respectively.

We next consider the case where local lattice operators only correspond to products of $N$ Ising order-parameter fields $\sigma_a$ or disorder-parameter fields $\mu_a$, such as $\sigma_1 \sigma_2 \sigma_3$ or $\sigma_1 \mu_2 \mu_3$ for $N=3$, in addition to thermal fields $\xi^a_R \xi^a_L$. 
In this case, the operator content of the field theory in Eq.~\eqref{eq:NIsingCFTHam} matches that of SO($N$)$_1$ WZW CFT. 
The interpretation of the ordered and disordered phases driven by $\lambda_1$ and positive $m$ depends on $N$ and also how local lattice operators are identified with the primary fields on the SO($N$)$_1$ WZW CFT. 
As we will discuss in more detail for specific examples in Sec.~\ref{sec:SPTPhaseTransitions}, these phases correspond to SPT phases protected by SO($N$) symmetry with either spontaneous or explicit dimerization accompanied by certain lattice symmetry breaking.
The universality class of the continuous phase transition is still the Ising one for $N=2$, whereas it becomes that of the SU(2)$_1$ WZW CFT for $N=3$.

\section{Coupled Ising chains}
\label{sec:CoupledIsing}

In this section, we consider a first lattice model in terms of $N$ coupled Ising chains, whose continuum limit corresponds to the field theory in Eq.~\eqref{eq:NIsingCFTHam}. 
This identification enables us to investigate numerically the nature of the phase transition for $N=2,3,4$.

As discussed in Refs.~\cite{Fabrizio00, YJWang00} for $N=2$ and $3$, the Hamiltonian in Eq.~\eqref{eq:NIsingCFTHam} has a lattice realization as given by
\begin{align} \label{eq:NIsingHam}
H_N &= \sum_{j=1}^L \left[ -\sum_{a=1}^N (J\sigma^z_{j,a} \sigma^z_{j+1,a} +h \sigma^x_{j,a}) -g \prod_{a=1}^N \sigma^z_{j,a} \right. \nonumber \\
&\quad \left. +K \sum_{a<b} (\sigma^z_{j,a} \sigma^z_{j+1,a} \sigma^z_{j,b} \sigma^z_{j+1,b} +\sigma^x_{j,a} \sigma^x_{j,b}) \right],
\end{align}
where $\sigma^{x,y,z}_{j,a}$ are Pauli matrices acting on $j$th site of the $a$th leg. 
Here, the first sum represents $N$ copies of the transverse field Ising chain, the second sum is a magnetic coupling among the chains, and the third sum is a coupling preserving the Kramers-Wannier self duality $\sigma^z_{j,a} \sigma^z_{j+1,a} \leftrightarrow \sigma^x_{j,a}$.
When the Ising chains are decoupled ($g=K=0$), there is an Ising critical point at $J=h$ separating an ordered phase with $\langle \sigma^z_{j,a} \rangle \neq 0$ ($J>h$) from a disordered phase with $\langle \sigma^z_{j,a} \rangle = 0$ ($J<h$).
In the vicinity of the (decoupled) Ising critical point $|g|, |K| \ll J=h$, the model parameters are related to the coupling constants in the CFT Hamiltonian, Eq.~\eqref{eq:NIsingCFTHam}, via $m \propto h-J$, $\lambda_1 \propto -g$, and $\lambda_2 \propto K$.
The phase transition we are interested in then happens between the disordered phase driven by a strong $h$ and an ordered phase driven by a strong $g$.

We note that when $K \neq 0$ and $g=0$, the model is known as an $N$-chain (or $N$-color) generalization \cite{Grest81, Fradkin84, Shankar85, Ceccatto91} of the quantum Ashkin-Teller model \cite{Ashkin43, Kohmoto81}. 
For $N=2$, the order-disorder transition has critical exponents continuously varying with $K$ as the model is dual to the spin-$1/2$ XXZ chain.
For $N \geq 3$, the phase transition still belongs to ($N$ copies of) the Ising transition for $K>0$ whereas it becomes a first-order transition for $K<0$.
While the sign of $K$ causes a striking difference in nature of the phase transition in the absence of $g$ (i.e., for the $N$-chain Ashkin-Teller model), it will be immaterial once we include a nonvanishing $g$; as suggested by the perturbative RG analysis in Sec.~\ref{sec:RG}, the coupling $K$ will flow to a negative value under the RG transformation whatever its initial value is.

We also note that the model \eqref{eq:NIsingHam} for $N=2$ and $K=0$ has been studied in Refs.~\cite{Xavier14, Ramos20} by density-matrix renormalization group (DMRG) and truncated conformal space approach, where a continuous phase transition described by Ising CFT has been identified.
The $N=2$ model with inequivalent couplings on each chain also has been studied in Refs.~\cite{Franchi23, Rossini26} from different perspectives.

Below, we first discuss that the lattice model in Eq.~\eqref{eq:NIsingHam} actually has a hidden SO($N$) symmetry, as anticipated from the corresponding field theory in Eq.~\eqref{eq:NIsingCFTHam}.
We then move to a strong-coupling analysis of Eq.~\eqref{eq:NIsingHam} in the limit of $g \to \infty$, which reveals that the $N=2,3$ cases have a continuous phase transition, as originally shown in Refs.~\cite{Fabrizio00, YJWang00}.
On the other hand, much less is known for a weak-coupling regime and also for $N \geq 4$.
We thus perform a numerical analysis using MPS to show that the phase transition is indeed a continuous phase transition for $N=2,3$ beyond the strong-coupling limit and becomes a first-order transition for $N=4$.

\subsection{Symmetry of the model}
\label{sec:NIsingSym}

In the absence of $g$, the model obviously has a $\mathbb{Z}_2^N$ symmetry generated by $\mathcal{G}_a = \prod_{j=1}^L \sigma^x_{j,a}$ and an $S_N$ symmetry permuting the chains. 
A nonzero $g$ reduces the former to its $\mathbb{Z}_2^{N-1}$ subgroup generated by $\{ \mathcal{G}_1 \mathcal{G}_2, \mathcal{G}_1 \mathcal{G}_3, \cdots, \mathcal{G}_1 \mathcal{G}_N \}$.
We here argue that the model actually has a larger symmetry, SO($N$), as anticipated from its low-energy effective theory in Eq.~\eqref{eq:NIsingCFTHam}, under the \emph{open boundary condition}.
Our argument is basically in line with Ref.~\cite{YJWang00} for $N=3$.

Let us perform the Jordan-Wigner transformation for each chain, 
\begin{align}
\sigma^z_{j, a} &= \kappa_a \left( \prod_{k=1}^{j-1} i \eta^{a}_{2k-1} \eta^a_{2k} \right) \eta^a_{2j-1}, \\
\sigma^x_{j,a} &= i\eta^{a}_{2j-1} \eta^a_{2j}.
\end{align}
Here, $\eta^a_\ell$ are Majorana fermion operators obeying the anticommutation relation $\{ \eta^a_\ell, \eta^b_{\ell'} \} = 2\delta_{ab} \delta_{\ell \ell'}$, and $\kappa_a$ are also anticommuting variables called Klein factors, which ensure the correct algebra of the Pauli operators, and obey $\{ \kappa_a, \kappa_b \} = 2\delta_{ab}$ and $[\kappa_a, \eta^b_\ell] = 0$.
Under open boundary conditions, the Hamiltonian in Eq.~\eqref{eq:NIsingHam} becomes
\begin{align} \label{eq:NMajoranaHam}
H_N &= \sum_{j=1}^L \left[ -iJ \bm{\eta}_{2j} \cdot \bm{\eta}_{2j+1} -ih \bm{\eta}_{2j-1} \cdot \bm{\eta}_{2j} \right. \nonumber \\
&\quad +g \epsilon_{N,j} \kappa_1 \kappa_2 \cdots \kappa_N \prod_{\ell=1}^{2j-1} \eta^1_\ell \eta^2_\ell \cdots \eta^N_\ell \nonumber \\
&\quad \left. +\frac{K}{2} \left( (i\bm{\eta}_{2j} \cdot \bm{\eta}_{2j+1})^2 + (i\bm{\eta}_{2j-1} \cdot \bm{\eta}_{2j})^2 +2N \right) \right],
\end{align}
where we have introduced $\bm{\eta}_\ell = (\eta^1_\ell, \eta^2_\ell, \cdots, \eta^N_\ell)^T$ and $\epsilon_{N,j} = (-1)^{N(N-1)(j-1)(2j-1)/2} \cdot i^{N(j-1)}$.

Let us consider a rotation $\bm{\eta}_\ell \to \bm{\eta}'_\ell = O \bm{\eta}_\ell$ by $O \in$  O($N$).
The SO($N$) subgroup of O($N$) is generated by the set of $N(N-1)/2$ operators $J_{ab}$,
\begin{align} \label{eq:SONGenerators}
J_{ab} = \sum_{\ell=1}^{2L} \frac{i}{2} \eta^a_\ell \eta^b_\ell,
\end{align}
which transforms $\eta^a_\ell$ as 
\begin{align}
e^{-i\theta J_{ab}} \eta^c_\ell e^{i\theta J_{ab}} = \begin{cases} 
\eta^a_\ell \cos \theta -\eta^b_\ell \sin \theta & (c=a) \\
\eta^a_\ell \sin \theta + \eta^b_\ell \cos \theta & (c=b) \\
\eta^c_\ell & \textrm{(otherwise)}
\end{cases}.
\end{align}
The parity operators $P_a = \prod_{j=1}^L i\eta^a_{2j-1} \eta^a_{2j}$, which satisfy $P_a^2 =I$, give reflections, 
\begin{align}
P_a^{-1} \eta^b_\ell P_a = \begin{cases} -\eta^b_\ell & (b=a) \\ \eta^b_a & (b \neq a) \end{cases}.
\end{align}
Thus, for any element $O$ of O($N$), we can find a unitary operator $U$ such that $U^{-1} \eta^a_\ell U = \sum_b O_{ab} \eta^b_\ell$ by suitably combining $e^{i\theta J_{ab}}$ and $P_a$.

It is then easy to find that $\bm{\eta}_\ell \cdot \bm{\eta}_{\ell'}$ are invariant under O$(N)$, and thus the Hamiltonian in Eq.~\eqref{eq:NMajoranaHam} has an O($N$) symmetry when $g=0$.
On the other hand, the product $\eta^1_\ell \eta^2_\ell \cdots \eta^N_\ell$ transforms as a pseudoscalar under O$(N)$, i.e., $\eta^1_\ell \eta^2_\ell \cdots \eta^N_\ell \to \det(O) \eta^1_\ell \eta^2_\ell \cdots \eta^N_\ell$ for $\bm{\eta}_\ell \to O \bm{\eta}_\ell$.
As the Hamiltonian in Eq.~\eqref{eq:NMajoranaHam} involves $2L-1$ products of $\eta^1_\ell \eta^2_\ell \cdots \eta^N_\ell$, it only has an SO($N$) symmetry when $g \neq 0$.
Since the generators of SO($N$) are sums of local operators in term of the Majorana fermions, any element of SO($N$) acts as an onsite symmetry in the Hamiltonian given by Eq.~\eqref{eq:NMajoranaHam}.
However, they take nonlocal forms in terms of the original spin operators due to the Jordan-Wigner strings. 
Thus, an element of SO($N$) generally acts as a non-onsite symmetry in the Hamiltonian given by Eq.~\eqref{eq:NIsingHam}, which is written in terms of the original spins, and only its $\mathbb{Z}_2^{N-1}$ subgroup generated by $\{ P_a P_b \}$ acts as an onsite symmetry.

Therefore, we conclude that the Hamiltonian in Eq.~\eqref{eq:NIsingHam} has an SO($N$) symmetry under open boundary conditions, which is generated by nonlocal unitary transformations, except its $\mathbb{Z}_2^{N-1}$ subgroup.
Under periodic boundary conditions, the Hamiltonian in Eq.~\eqref{eq:NIsingHam} has a more complicated symmetry structure; depending of the sector of $\mathbb{Z}_2^{N-1}$ symmetry, different subgroups of SO($N$) leave the Hamiltonian invariant.

\subsection{Strong-coupling Hamiltonians}

In order to derive the strong-coupling Hamiltonian for Eq.~\eqref{eq:NIsingHam} in the limit $g \to \infty$, we first consider the following unitary transformation
\begin{align}
U_N = \prod_{j=1}^L \prod_{a=1}^{N-1} \textrm{CNOT}_{(j, a), (j,N)},
\end{align}
where $\textrm{CNOT}_{(i,a),(j,b)}$ is the controlled-NOT (or controlled-X) operation defined by 
\begin{align}
\textrm{CNOT}_{(i,a),(j,b)} = e^{i\pi (1-\sigma^z_{i,a})(1-\sigma^x_{j,b})/4}
\end{align}
and transforms the Pauli operators at the sites $(i,a)$ and $(j,b)$ as 
\begin{align}
\begin{split}
\sigma^x_{i,a} &\mapsto \sigma^x_{i,a} \sigma^x_{j,b}, \\
\sigma^x_{j,b} &\mapsto \sigma^x_{j,b}, \\
\sigma^z_{i,a} &\mapsto \sigma^z_{i,a}, \\
\sigma^z_{j,b} &\mapsto \sigma^z_{i,a} \sigma^z_{j,b}.
\end{split}
\end{align}
Under this unitary transformation, the Hamiltonian in Eq.~\eqref{eq:NIsingHam} is transformed into $\tilde{H}_N = U_N^{-1} H_N U_N$, where $\tilde{H}_N$ is given by
\begin{align}
\tilde{H}_N 
&= \sum_{j=1}^L \Biggl[ -\sum_{a=1}^{N-1} (J \sigma^z_{j,a} \sigma^z_{j+1,a} +h \sigma^x_{j,a} \sigma^x_{j,N}) \nonumber \\
&\quad -J \sigma^z_{j,N} \sigma^z_{j+1,N}\prod_{a=1}^{N-1} \sigma^z_{j,a} \sigma^z_{j+1,a}  -h \sigma^x_{j,N} -g \sigma^z_{j,N} \nonumber \\
&\quad +K \sum_{a<b<N} (\sigma^z_{j,a} \sigma^z_{j+1,a} \sigma^z_{j,b} \sigma^z_{j+1,b} +\sigma^x_{j,a} \sigma^x_{j,b}) \nonumber \\
&\quad +K \sum_{a=1}^{N-1} \Biggl( \sigma^z_{j,N} \sigma^z_{j+1,N} \prod_{\substack{b \neq a \\ b<N}} \sigma^z_{j,b} \sigma^z_{j+1,b} +\sigma^x_{j,a} \Biggr) \Biggr]
\end{align}
for $N \geq 3$ and by
\begin{align}
\tilde{H}_2
&= \sum_{j=1}^L \bigl[ -J \sigma^z_{j,1} \sigma^z_{j+1,1} (1+\sigma^z_{j,2} \sigma^z_{j+1,2}) -h (\sigma^x_{j,1}+1) \sigma^x_{j,2} \nonumber \\
&\quad -g \sigma^z_{j,2} +K (\sigma^z_{j,2} \sigma^z_{j+1,2} + \sigma^x_{j,1}) \bigr]
\end{align}
for $N=2$.

In the strong-coupling limit $g \to \infty$, $\sigma^z_{j,N}$ is pinned and takes a finite expectation value $\langle \sigma^z_{j,N} \rangle = 1$.
We then perform degenerate perturbation theory for $2^{L(N-1)}$-fold degenerate ground states spanned by $\sigma^\alpha_{j,a}$ with $a=1,\cdots,N-1$ to find an effective Hamiltonian up to the second order, 
\begin{align}
\tilde{H}^\textrm{eff}_N &= \sum_{j=1}^L \Biggl[-J\sum_{a=1}^{N-1} \sigma^z_{j,a} \sigma^z_{j+1,a} +\left( K-\frac{h^2}{g} \right) \sum_{a=1}^{N-1} \sigma^x_{j,a} \nonumber \\
&\quad +K \sum_{a<b<N} \sigma^z_{j,a} \sigma^z_{j+1,a} \sigma^z_{j,b} \sigma^z_{j+1,b} \nonumber \\
&\quad +\left( K-\frac{h^2}{g} \right) \sum_{a<b<N} \sigma^x_{j,a} \sigma^x_{j,b} -J \prod_{a=1}^{N-1} \sigma^z_{j,a} \sigma^z_{j+1,a} \nonumber \\
&\quad +K\sum_{a=1}^{N-1} \prod_{\substack{b \neq a \\ b<N}} \sigma^z_{j,b} \sigma^z_{j+1,b} \Biggr]
\end{align}
for $N \geq 3$ and
\begin{align}
\tilde{H}^\textrm{eff}_2 = \sum_{j=1}^L \left[ -2J \sigma^z_{j,1} \sigma^z_{j+1,1} +\left( K-\frac{h^2}{g} \right) \sigma^x_{j,1} \right]
\end{align}
for $N=2$.
Here, we have neglected additive constants.

When $N=2$, the effective Hamiltonian reduces to a single chain of the transverse field Ising chain \cite{Fabrizio00}, so that the Ising transition between the ordered and disordered phases is expected at $2J \sim K-h^2/g$.
When $N=3$, the effective Hamiltonian reduces to the quantum Ashkin-Teller model \cite{YJWang00},
\begin{align}
\tilde{H}^\textrm{eff}_3& = \sum_{j=1}^L \biggl[ (K-J) (\sigma^z_{j,1} \sigma^z_{j+1,1} +\sigma^z_{j,2} \sigma^z_{j+1,2}  \nonumber \\
&\quad + \sigma^z_{j,1} \sigma^z_{j+1,1} \sigma^z_{j,2} \sigma^z_{j+1,2}) \nonumber \\
&\quad +\left( K-\frac{h^2}{g} \right) (\sigma^x_{j,1} +\sigma^x_{j,2} +\sigma^x_{j,1} \sigma^x_{j,2}) \biggr].
\end{align}
Therefore, the phase transition is expected to occur at $J \sim h^2/g$ and is described by the four-state Potts CFT with central charge $c=1$.
It has been shown that this model is dual to the dimerized Heisenberg chain \cite{Kohmoto81, Kohmoto92},
\begin{align}
\label{eq:DualHeisenbergChain}
\tilde{H}_3^\textrm{eff} &= \sum_{j=1}^L \biggl[ (J-K)(\tau^x_{2j-1} \tau^x_{2j} +\tau^y_{2j-1} \tau^y_{2j} +\tau^z_{2j-1} \tau^z_{2j}) + \nonumber \\
&\quad + \left( \frac{h^2}{g} -K \right) (\tau^x_{2j} \tau^x_{2j+1} +\tau^y_{2j} \tau^y_{2j+1} +\tau^z_{2j} \tau^z_{2j+1}) \biggr],
\end{align}
where $\tau^\alpha_l$ are Pauli operators. The phase transition for the latter model is described by the SU(2)$_1$ WZW CFT.

For $N \geq 4$, on the other hand, the effective Hamiltonian cannot be reduced to any known model to our knowledge.
For example, the effective Hamiltonian for $N=4$ is given by
\begin{align}
\tilde{H}^\textrm{eff}_4 &= \sum_{j=1}^L \biggl[ -J \sum_{a=1}^3 \sigma^z_{j,a} \sigma^z_{j+1,a} +\left( K-\frac{h^2}{g} \right) \sum_{a=1}^3 \sigma^x_{j,a} \nonumber \\
&\quad +2K \sum_{a<b} \sigma^z_{j,a} \sigma^z_{j+1,a} \sigma^z_{j,b} \sigma^z_{j+1,b} \nonumber \\
&\quad +\left( K-\frac{h^2}{g} \right) \sum_{a<b} \sigma^x_{j,a} \sigma^x_{j,b} -J \prod_{a=1}^3 \sigma^z_{j,a} \sigma^z_{j+1,a} \biggr],
\end{align}
which can be regarded as the three-chain generalization of the Ashkin-Teller model with an additional six-spin interaction. 
Since the six-spin interaction is of the order of $J$, its effect on the transition cannot be addressed by perturbative approaches.

\subsection{Numerical results}
\label{sec:IsingNum}

In this section, we set $J=1$.
For fixed values of $g$ and $K$, we vary $h$ and numerically investigate the nature of phase transitions from the ordered phase with $\langle \sigma^z_{j,a} \rangle \neq 0$ to the disordered phase. 
We employ infinite MPS methods with a fixed bond dimension $\chi$ to variationally obtain the ground states of the Hamiltonian in Eq.~\eqref{eq:NIsingHam} in the thermodynamic limit. 
We selectively use either infinite DMRG (iDMRG) \cite{McCulloch08} or variational uniform MPS (VUMPS) \cite{Zauner-Stauber18, Vanderstraeten19} algorithm, choosing the algorithm that achieves better convergence.

In order to identify the phase transition points, we compute the magnetization
\begin{align}
M \equiv \langle \sigma^z_{j,1} \rangle
\end{align}
using VUMPS by assuming $N$-site unit cell (i.e., each unit cell contains $N$ spin-$1/2$'s) for the infinite MPS and implementing no symmetry.
Here, we denote by $\langle X \rangle$ the ground-state expectation value of $X$.
The magnetization works as an order parameter to detect spontaneous breaking of $\mathbb{Z}_2^{N-1}$ spin-flip symmetry as discussed in Sec.~\ref{sec:NIsingSym}.

In order to accurately extract the ground-state properties near the transition point $h=h_c$, we compute the von Neumann entanglement entropy $S_\textrm{vN}$ under half cut of the chain, the correlation length $\xi$ extracted from the eigenvalues of an MPS transfer matrix, and various connected correlation functions $\langle O_i O_j \rangle_c \equiv \langle O_i O_j \rangle - \langle O_i \rangle \langle O_j \rangle$. 
For this purpose, we use iDMRG for $N=2$ with explicitly preserving $\mathbb{Z}_2 \times \mathbb{Z}^C_2$ symmetry related to the spin flip and chain permutation and assume a four-site unit cell, i.e., each unit cell containing $2 \times 2 = 4$ spin-$1/2$'s. 
For $N \geq 3$, we use VUMPS with explicitly preserving $\mathbb{Z}_2^{N-1}$ spin-flip symmetry and assume an $N$-site unit cell. 

We note that such symmetric MPS can efficiently describe the paramagnetic phase above the transition point, $h > h_c$. 
However, these symmetric MPS become less efficient than non-symmetric MPS in the magnetically-ordered phase below the transition point $h < h_c$, because of an additive entanglement caused by cat-state formation to respect the imposed symmetry.
Thus, in principle, one should use different MPS to achieve better accuracy within each phase.
Nevertheless, we mainly use the symmetric MPS in the following analysis, since we are ultimately interested in the nature of the phase transitions and if it is continuous, the full symmetry is restored at the transition point.
The latter is expected for $N=2,3$.
For $N=4$, we expect a first-order transition for which symmetric MPS may not be the best choice. 
However, as shown in Appendix~\ref{app:Ising}, non-symmetric MPS also show the signatures of the first-order transition.
Therefore, the conclusion about the nature of the transition for $N=4$ will not be altered by the choice of symmetric or non-symmetric MPS.

\subsubsection{$N=2$}

\begin{figure}[tb]
\includegraphics[width=0.48\textwidth]{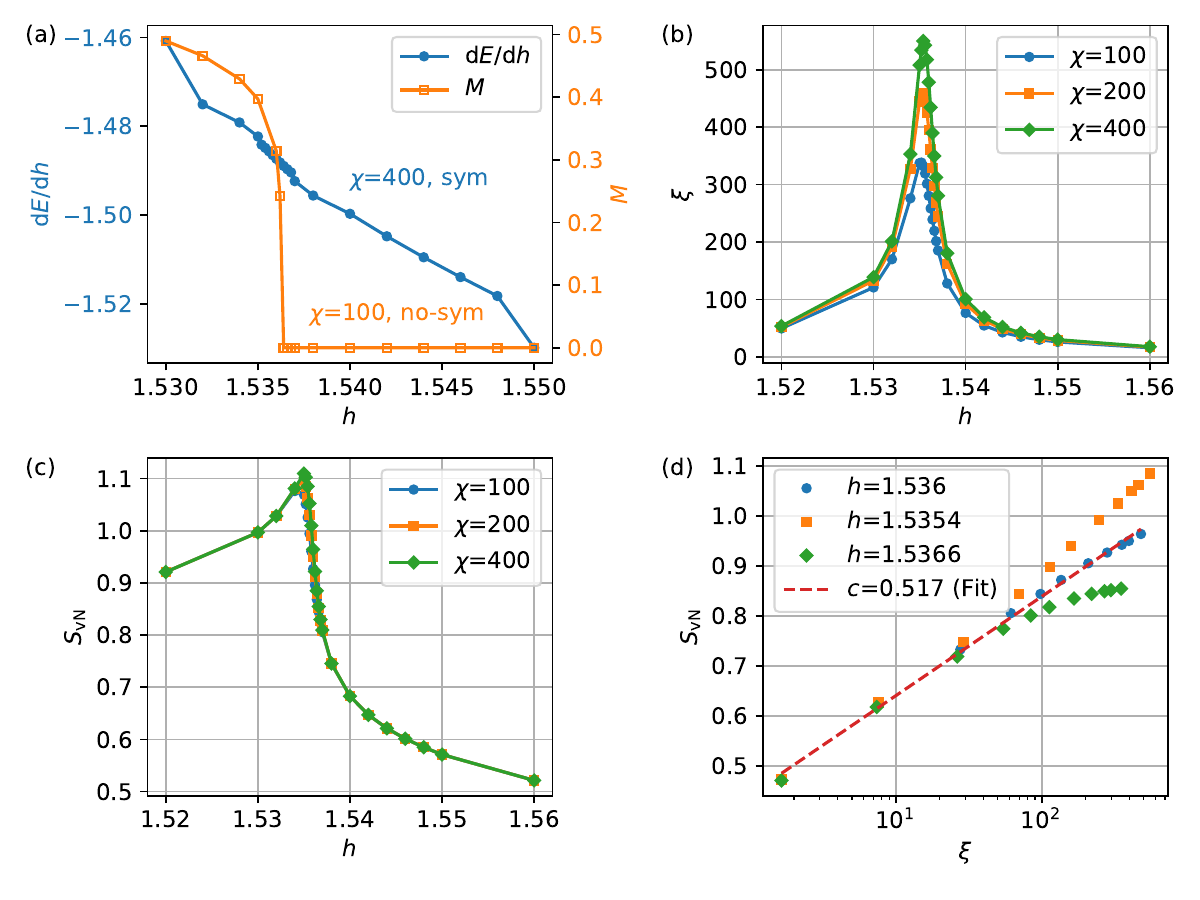}
\caption{Numerical results for the $N=2$ coupled Ising chains in Eq.~\eqref{eq:NIsingHam} with $K=0$ and $g=0.5$.
(a) Derivative of the ground-state energy density $dE/dh$ and m}agnetization $M$, (b) correlation length $\xi$, and (c) half-system von Neumann entanglement entropy $S_\textrm{vN}$ are plotted as functions of $h$. 
(d) Plot of $S_\textrm{vN}$ vs the correlation length $\xi$ in the vicinity of the transition. 
The dashed line is a fitting function of the form $(c/6) \ln \xi +c_0$.
\label{fig:Coupled2Ising}
\end{figure}

For $N=2$, we show our numerical results for $K=0$ and $g=0.5$ in Fig.~\ref{fig:Coupled2Ising}.
We first compute the magnetization $M$ using non-symmetric MPS with $\chi=100$, which takes a finite value for small $h$ as shown in Fig.~\ref{fig:Coupled2Ising}(a), indicating a magnetic long-range order.
The magnetization vanishes around the transition point $h \sim 1.536$, while the derivative of the ground-state energy density $dE/dh$ computed from symmetric MPS with $\chi=400$ changes smoothly. 
Furthermore, as shown in Figs.~\ref{fig:Coupled2Ising}(b) and (c), both correlation length $\xi$ and entanglement entropy $S_\textrm{vN}$ show a divergent behavior at the transition.
These signatures support the existence of a continuous phase transition beyond the strong-coupling limit.

To pin down the universality class of this phase transition, we examine the scaling form of the entanglement entropy $S_\textrm{vN}$ with the correlation length $\xi$ near the transition point $h \sim 1.536$. 
As shown in Fig.~\ref{fig:Coupled2Ising}(d), the entanglement entropy right at the transition $h=1.536$ can be fitted with a logarithmic function $(c/6)\ln\xi + c_0$ predicted by CFT, from which we can extract the central charge $c \sim 0.517$, in agreement with $c=1/2$ expected for the Ising CFT. 
It saturates to a constant value above the transition as a finite gap is expected in the disordered phase, whereas a drift towards a larger value is observed below the transition; the latter is due to an additive entanglement of $\ln 2$ coming from a cat-state formation in the ordered phase, since our MPS explicitly preserves the $\mathbb{Z}_2$ spin-flip symmetry, which is spontaneously broken in this phase.
As we will see below, similar drifts can also be found for $N=3$ and $4$ as we generally expect an additive entanglement of $(N-1)\ln 2$ in the ordered phase.

\begin{figure}[tb]
\includegraphics[width=0.48\textwidth]{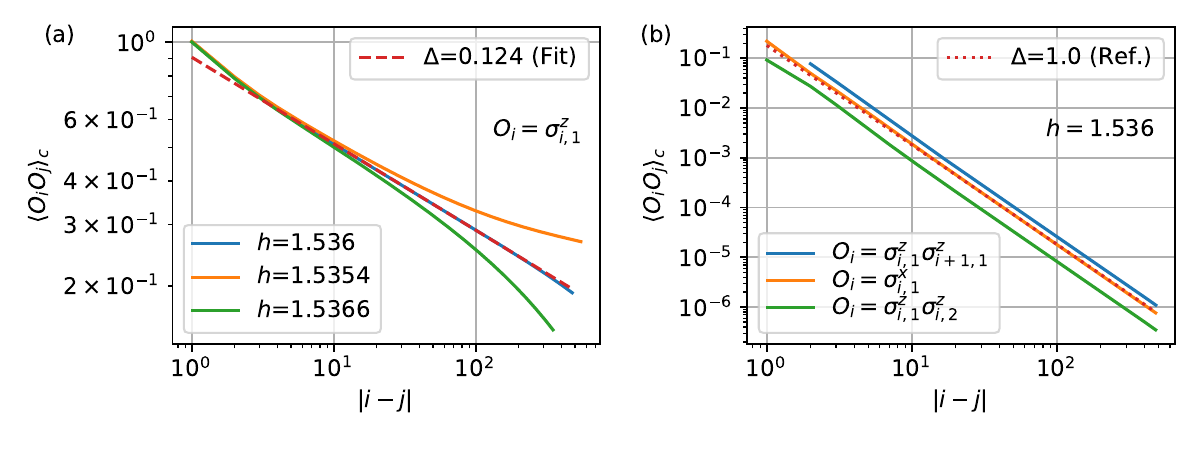}
\caption{Numerical results  for the $N=2$ coupled Ising chains in Eq.~\eqref{eq:NIsingHam} with $K=0$ and $g=0.5$. 
Two-point connected correlation functions $\langle O_i O_j \rangle_c$ for (a) $O_i = \sigma^z_{i,1}$ and (b) $O_i = \sigma^z_{i,1} \sigma^z_{i+1,1}$, $\sigma^x_{i,1}$, and $\sigma^z_{i,1} \sigma^z_{i,2}$ are computed in the vicinity of the phase transitions. 
The dashed line is a fitting function of the power-law form $C |i-j|^{-2\Delta}$ in (a), whereas it is of the form $C|i-j|^{-2}$ shown as a reference in (b).}
\label{fig:Coupled2IsingCC}
\end{figure}

In order to extract critical exponents at the transition $h=1.536$, we compute two-point connected correlation functions $\langle O_i O_j \rangle_c$ with $O_i = \sigma^z_{i,1}$, $\sigma^z_{i,1} \sigma^z_{i+1,1}$, $\sigma^x_{i,1}$, and $\sigma^z_{i,1} \sigma^z_{i,2}$, which are shown in Fig.~\ref{fig:Coupled2IsingCC}.
The correlation functions are well fitted into the power-law scaling form $C |i-j|^{-2\Delta}$ in the range $|i-j| \in [5,\xi/2]$, where $\Delta$ should be the scaling dimension of an associated field in the underlying CFT. 
We find $\Delta \sim 0.124$ for $O_i = \sigma^z_{i,1}$ and $\Delta \sim 1$ for the other three, which is in excellent agreement with the scaling dimension $\Delta_\sigma=1/8$ and $\Delta_\epsilon=1$ expected for the order-parameter and thermal field at the Ising transition, respectively.
These results are consistent with the previous numerical results for $K=0$ \cite{Xavier14, Ramos20}.
We have obtained quantitatively similar results on the critical behavior at transition points for other sets of the parameters $g$ and $K$, as shown in Fig.~\ref{fig:Coupled2IsingSuppl} of Appendix~\ref{app:Ising}. 
The extracted central charges and critical exponents at the transitions are summarized in Table~\ref{tab:Coupled2Ising}.
\begin{table}
\begin{ruledtabular}
\begin{tabular}{llllll}
$g$ & $K$ & $h_c$ & $c$ & $\Delta(\sigma^z_{i,1})$ & $\Delta(\sigma^x_{i,1})$ \\ \hline
0.5 & 0.0 & 1.5360 & 0.517 & 0.124 & 1.006 \\
2.0 & 0.0 & 2.2968 & 0.506 & 0.125 & 1.002 \\
0.5 & $-0.3$ & 1.4646 & 0.497 & 0.130 & 1.009 \\
2.0 & $-0.3$ & 2.1652 & 0.515 & 0.122 & 1.003 \\
0.5 & 0.3 & 1.6293 & 0.520 & 0.126 & 1.003 \\
2.0 & 0.3 & 2.4316 & 0.508 & 0.127 & 1.002 \\ \hline
\multicolumn{3}{l}{Ising (exact)} & 0.5 & 0.125 & 1
\end{tabular}
\end{ruledtabular}
\caption{Central charge $c$ and critical exponents $\Delta(O_i)$ for $O_i = \sigma^z_{i,1}$ and $\sigma^x_{i,1}$ numerically extracted at the transition points $h=h_c$ for the $N=2$ coupled Ising chains in Eq.~\eqref{eq:NIsingHam} for various $g$ and $K$. Exact results for the Ising CFT are also given.}
\label{tab:Coupled2Ising}
\end{table}
This confirms that the $N=2$ coupled Ising chains in Eq.~\eqref{eq:NIsingHam} has a continuous phase transition described by the Ising CFT with $c=1/2$.

\subsubsection{$N=3$}

\begin{figure}[tb]
\includegraphics[width=0.48\textwidth]{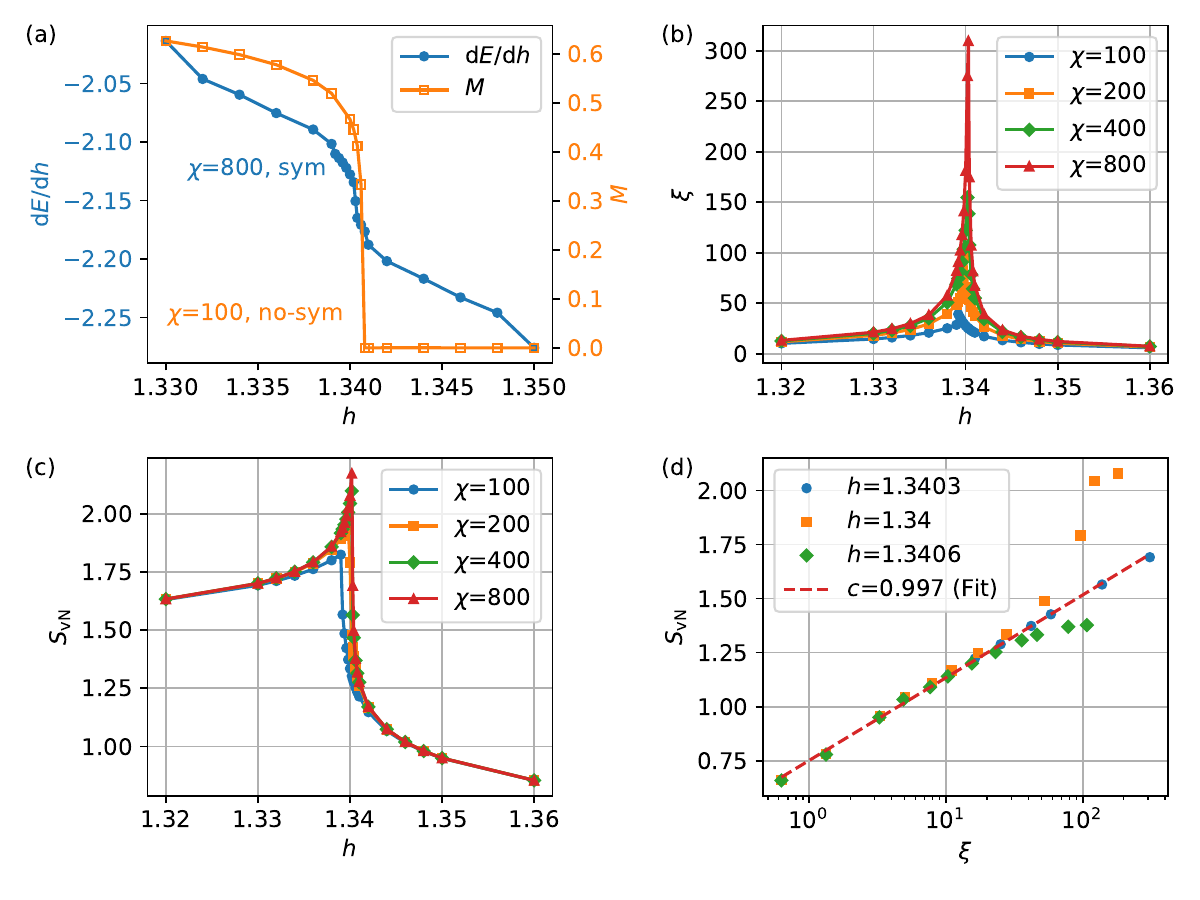}
\caption{Numerical results for the $N=3$ coupled Ising chains in Eq.~\eqref{eq:NIsingHam} with $K=0$ and $g=0.5$.
(a) Derivative of the ground-state energy density $dE/dh$ and magnetization $M$, (b) correlation length $\xi$, and (c) von Neumann entanglement entropy are plotted as functions of $h$. 
(d) $S_\textrm{vN}$ are plotted against the correlation length $\xi$ at the vicinity of the transition. 
The dashed line is a fitting function of the form $(c/6) \ln \xi +c_0$.}
\label{fig:Coupled3Ising}
\end{figure}

For $N=3$, we show our numerical results for $K=0$ and $g=0.5$ in Fig.~\ref{fig:Coupled3Ising}. 
Similarly to the $N=2$ case, the magnetization $M$ computed from non-symmetric MPS with $\chi=100$ vanishes around the transition point $h \sim 1.34$, while the derivative of the ground-state energy density $dE/dh$ computed from symmetric MPS with $\chi=800$ varies smoothly [Fig.~\ref{fig:Coupled3Ising}(a)].
At this transition point, both correlation length and entanglement entropy show a divergent behavior [Figs.~\ref{fig:Coupled3Ising}(b) and (c)]. 
As shown in Fig.~\ref{fig:Coupled3Ising}(d), the entanglement entropy $S_\textrm{vN}$ at the vicinity of the transition is fitted well to the CFT scaling form $(c/6) \ln \xi + c_0$ with $c \sim 0.997$. 
This suggests that the $N=3$ coupled Ising chains in Eq.~\eqref{eq:NIsingHam} has a continuous phase transition described by a $c=1$ CFT.

\begin{figure}[tb]
\includegraphics[width=0.48\textwidth]{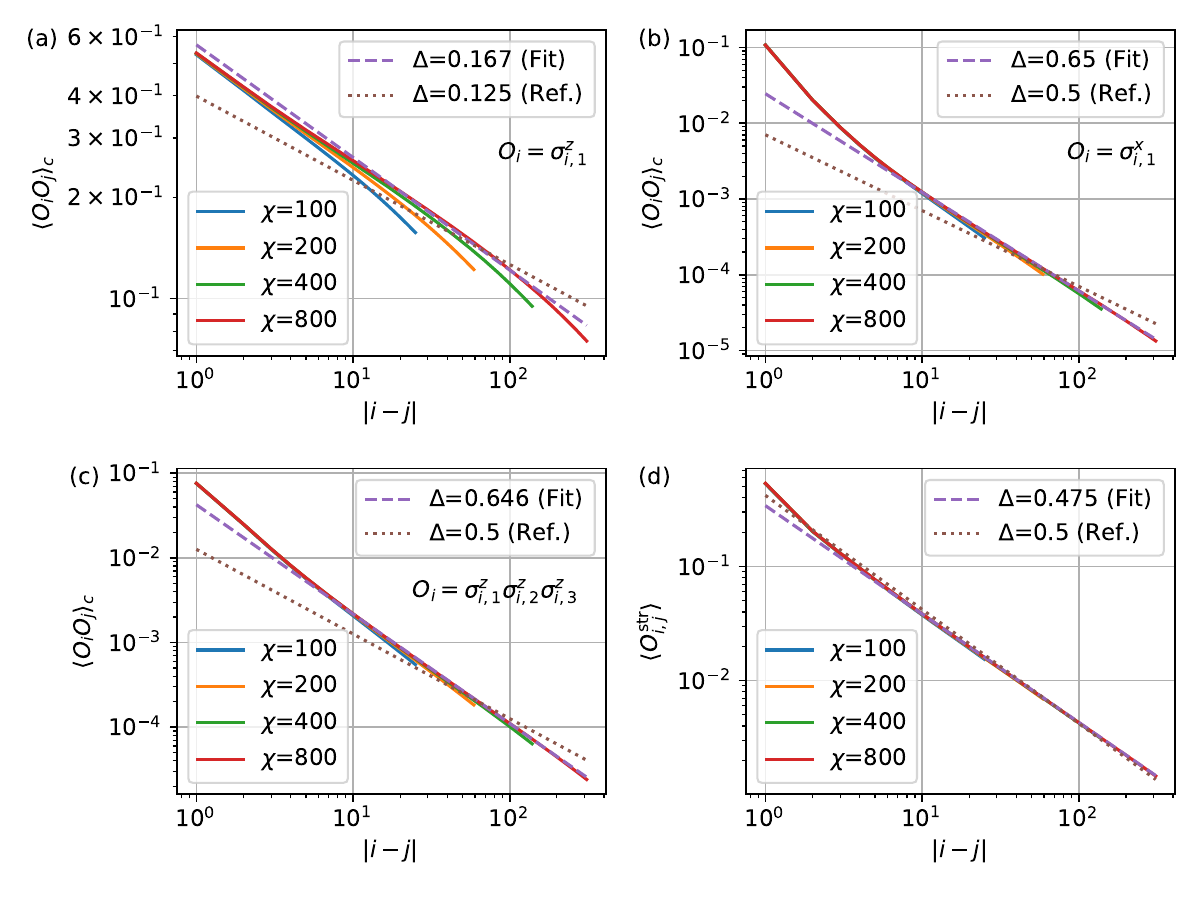}
\caption{Numerical results for the $N=3$ coupled Ising chains \eqref{eq:NIsingHam} with $K=0$ and $g=0.5$ at the transition $h=1.3403$.
Two-point connected correlation functions $\langle O_i O_j \rangle_c$ for (a) $O_i = \sigma^z_{i,1}$, (b) $O_i = \sigma^z_{i,1}$, and (c) $O_i = \sigma^z_{i,1} \sigma^z_{i,2} \sigma^z_{i,3}$ are computed. 
The string correlation function $\langle O^\textrm{str}_{i,j} \rangle$ is also shown in (d).
The blue dashed lines are fitting functions of the power-law form $C |i-j|^{-2\Delta}$, whereas the orange ones are those with exact exponents of the four-state Potts CFT shown as references.}
\label{fig:Coupled3IsingCC}
\end{figure}

While the strong-coupling analysis indicates that the corresponding CFT is the four-state Potts CFT, it does not necessarily persist for small $g$ as the $c=1$ CFT can have continuously varying critical exponents. 
We thus compute the connected correlation functions $\langle O_i O_j \rangle_c$ for $O_i = \sigma^z_{i,1}$, $\sigma^x_{i,1}$, and $\sigma^z_{i,1} \sigma^z_{i,2} \sigma^z_{i,3}$ in the vicinity of the transition, which are shown in Figs.~\ref{fig:Coupled3IsingCC}(a)--(c), respectively.
Their scaling behaviors largely deviate from the expected behaviors for the four-state Potts CFT \cite{Kohmoto81, SKYang87}.
For $O_i = \sigma^z_{i,1}$, we expect the scaling dimension $\Delta_\sigma = 1/8$ of the order-parameter field but the fitting analysis yields $\Delta \sim 0.167$. 
For $O_i = \sigma^x_{i,1}$ and $\sigma^z_{i,1} \sigma^z_{i,2} \sigma^z_{i,3}$, which correspond to the thermal field in the four-state Potts CFT and the dimerization order parameter field in the SU(2)$_1$ WZW CFT, respectively, we expect $\Delta_\epsilon=0.5$ but we actually obtain $\Delta \sim 0.65$ by fitting.

One may think that these strong deviations from the expected critical exponents preclude the possibility of the four-state Potts CFT at the continuous phase transition.
However, unlike the $N=2$ case governed by the Ising CFT, it has been known that the four-state Potts CFT has marginally irrelevant operators as symmetry-allowed perturbations. 
These operators cause logarithmic corrections to finite-size quantities and make it difficult to extract their true asymptotic behaviors \cite{Cardy86}. 
In fact, previous works on the four-state Potts criticality found strong deviations from expected scaling behaviors for finite-size excitation gaps \cite{Alcaraz87} or correlation functions (to be more precise, the string correlators of the dual Heisenberg chain) \cite{Qin03, Lou03}.
In order to assess these finite-size corrections, we directly apply our MPS analysis to the quantum Ashkin-Teller model at the four-state Potts criticality, which corresponds to Eq.~\eqref{eq:NIsingHam} with $N=2$, $J=h=-K$, and $g=0$. 
As shown in Appendix~\ref{app:AT}, the connected correlation functions corresponding to the order-parameter and thermal fields, obtained with $\chi=800$, yield the critical exponents $\Delta \sim 0.162$ and $0.61$, respectively.
These values are reasonably close to those obtained for the $N=3$ coupled Ising chains with $K=0$ and $g=0.5$ and also with the other choices of $K$ and $g$, as shown in Appendix~\ref{app:Ising} and summarized in Table~\ref{tab:Coupled3Ising}.
\begin{table}
\begin{ruledtabular}
\begin{tabular}{lllllll}
$g$ & $K$ & $h_c$ & $c$ & $\Delta(\sigma^z_{i,1})$ & $\Delta(\sigma^x_{i,1})$ & $\Delta(O^\textrm{str}_{i,j})$ \\ \hline
0.5 & 0.0 & 1.34030 & 0.997 & 0.167 & 0.650 & 0.475 \\
2.0 & 0.0 & 1.85038 & 1.030 & 0.154 & 0.620 & 0.476 \\
0.5 & $-0.3$ & 1.28690 & 1.039 & 0.161 & 0.599 & 0.486 \\
2.0 & $-0.3$ & 1.78780 & 1.020 & 0.166 & 0.614 & 0.481 \\
0.5 & 0.3 & 1.43374 & 0.959 & 0.164 & 0.719 & 0.467 \\
2.0 & 0.3 & 1.94036 & 1.009 & 0.159 & 0.643 & 0.472 \\ \hline
\multicolumn{3}{l}{Ashkin-Teller (num.)} & N.A. & 0.162 & 0.614 & 0.478 \\
\multicolumn{3}{l}{Four-state Potts (exact)} & 1 & 0.125 & 0.5 & 0.5
\end{tabular}
\end{ruledtabular}
\caption{Central charge $c$ and critical exponents $\Delta(O_i)$ for the correlation functions with $O_i = \sigma^z_{i,1}$ and $\sigma^x_{i,1}$ and for the string correlation function $\langle O^\textrm{str}_{i,j} \rangle$ numerically extracted at the transition points $h=h_c$ for the $N=3$ coupled Ising chains in Eq.~\eqref{eq:NIsingHam} for various $g$ and $K$.
Numerical results for the Ashkin-Teller model at the four-state Potts criticality and exact results for the four-state Potts CFT are also shown.}
\label{tab:Coupled3Ising}
\end{table}

In addition to the correlation functions of local operators, we also compute the string correlation function
\begin{align}
\langle O^\textrm{str}_{i,j} \rangle = \left< \sigma^z_{i,1} \left( \prod_{k=i+1}^{j-1} \sigma^x_{k,2} \sigma^x_{k,3} \right) \sigma^z_{j,1} \right>,
\end{align}
which corresponds to the spin-spin correlation function $\langle \tau^x_{2i} \tau^x_{2j} \rangle$ for the dual Heisenberg chain in Eq.~\eqref{eq:DualHeisenbergChain} in the strong-coupling limit \cite{Kohmoto81, YJWang00}. 
We thus expect a power-law scaling with the exponent $\Delta_g=1/2$, which corresponds to the scaling dimension for the primary field of the SU(2)$_1$ WZW CFT.
As shown in Fig.~\ref{fig:Coupled3IsingCC}(d), the numerical results are fitted well with a power-law function with $\Delta \sim 0.475$.
We also obtain the exponent $\Delta \sim 0.478$ from the numerical simulation of the Ashkin-Teller model at the four-state Potts criticality (see Appendix~\ref{app:AT}).
These values, including those obtained for other choices of $K$ and $g$ in the $N=3$ model, are relatively close to the expected exponent $\Delta_g=1/2$, compared to the correlation functions of local operators discussed above.

If we suppose that the continuous phase transition is described by the four-state Potts CFT, we could tune marginal operators by appropriately deforming the $N=3$ coupled Ising Hamiltonian in Eq.~\eqref{eq:NIsingHam}.
For simplicity, let us assume $K=0$.
As discussed in Sec.~\ref{sec:NIsingSym}, the Hamiltonian possesses an SO(3) symmetry under the open boundary condition. 
This SO(3) symmetry can be broken down to SO(2) $\sim$ U(1) when we vary the coupling constants $J$ and/or $h$ in one of the three chains.
This would correspond to tuning of one of the three marginal operators in the four-state Potts CFT \cite{Dijkgraaf89, Thorngren24} and lead to a continuous family of the $c=1$ free-boson CFT at the phase transition, akin to the Ashkin-Teller model or XXZ chain.
For example, setting $J=h=0$ in one chain obviously yields two copies of the Ising chain, which has a critical point described by a massless Dirac CFT with $\Delta_\sigma=1/8$ and $\Delta_\epsilon=1$, irrespective of the value of $g$.
One could also tune all three marginal operators without altering the four-state Potts CFT by adding appropriate spin interactions corresponding to the next-nearest-neighbor Majorana couplings $\bm{\eta}_\ell \cdot \bm{\eta}_{\ell+2}$.
This might eliminate the marginally irrelevant coupling in the four-state Potts CFT and allow us to numerically extract its critical exponents without facing logarithmic corrections (see, e.g., Ref.~\cite{Okamoto92}).

To sum up, we numerically find a continuous phase transition for the $N=3$ coupled Ising chains beyond the strong-coupling limit.
Given that the same MPS approach with the same order of bond dimensions yields quite similar critical exponents for the correlation functions in both $N=3$ models and Ashkin-Teller model at the four-state Potts criticality, we expect that the continuous phase transition for $N=3$ belongs to the universality class of the four-state Potts CFT.
However, it seems to require more numerical effort to unambiguously identify the universality class of this continuous phase transition, which is left for future work.

\subsubsection{$N=4$}

\begin{figure}[tb]
\includegraphics[width=0.48\textwidth]{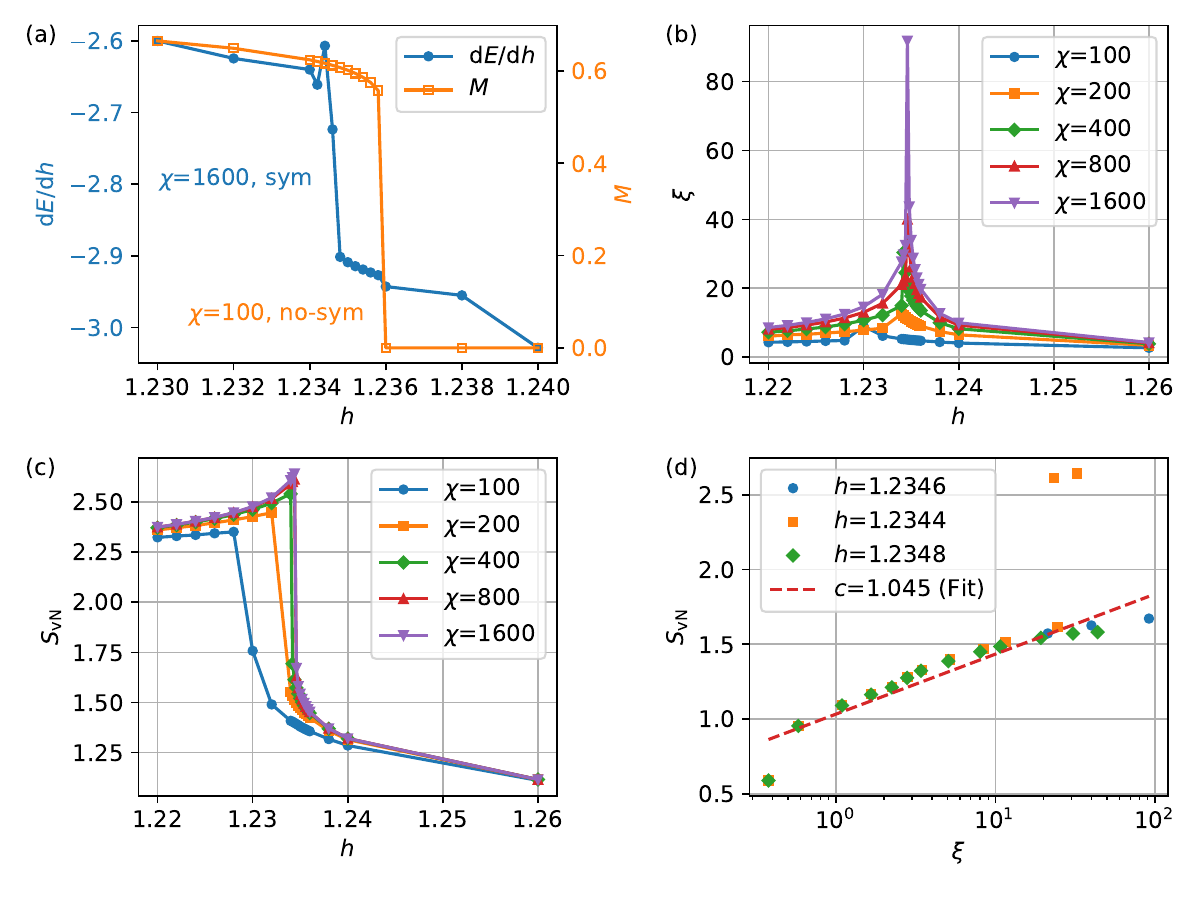}
\caption{Numerical results for the $N=4$ coupled Ising chains in Eq.~\eqref{eq:NIsingHam} with $K=0$ and $g=0.5$.
(a) Derivative of the ground-state energy density $dE/dh$ and magnetization $M$, (b) correlation length $\xi$, and (c) half-system von Neumann entanglement entropy $S_\textrm{vN}$ are plotted as functions of $h$. 
(d) Plot of $S_\textrm{vN}$ vs the correlation length $\xi$ in the vicinity of the transition.}
\label{fig:Coupled4Ising}
\end{figure}

For $N=4$, we show our numerical results for $K=0$ and $g=0.5$ in Fig.~\ref{fig:Coupled4Ising}.
As shown in Fig.~\ref{fig:Coupled4Ising}(a), the magnetization computed from non-symmetric MPS with $\chi=100$ vanishes around the transition point $h \sim 1.235$, but its change is rather abrupt compared with the $N=2$ and $N=3$ cases. 
Indeed, the derivative of the ground-state energy density $dE/dh$ computed from symmetric MPS with $\chi=1600$ shows a discontinuous change near the transition.
Since we are directly treating the thermodynamic limit by using infinite MPS, the discontinuity in the first-order derivative of the ground-state energy density is a defining feature of the first-order transition.
Physically, this transition is driven by a level crossing between two macroscopically distinct gapped ground states, corresponding to the ordered and disordered phases.

To further substantiate this claim, we examine the gapped nature of the transition through the entanglement entropy and correlation length.
As shown in Figs.~\ref{fig:Coupled4Ising}(b)-(c), both entanglement entropy and correlation length show a peak at the transition, but the entanglement entropy actually shows a saturating behavior against the correlation length.
As shown in Fig.~\ref{fig:Coupled4Ising}(d), slightly inside the ordered phase $h < 1.2346$, the entanglement entropy grows in the logarithmic form for small bond dimensions but suddenly jumps to a higher value for large bond dimensions, leading to a discontinuous behavior as a function of the correlation length $\xi$. 
This behavior is a direct numerical manifestation of the aforementioned level crossing, where the variational MPS switches from one ground-state branch to another upon increasing the bond dimension.
Right at or above the transition $h \geq 1.2346$, the entanglement entropy shows sublogarithmic scaling behaviors and appears to saturate to constant values for large bond dimensions, indicating the presence of a finite excitation gap even at the transition.
As shown in Appendix~\ref{app:Ising}, similar signatures of the first-order transition have also been found for other parameters of $K$ and $g$ and also using non-symmetric MPS. 
These numerical results strongly indicate that the $N=4$ coupled Ising chains in Eq.~\eqref{eq:NIsingHam} have a first-order transition.

\section{Phase transitions in SO($N$)-symmetric spin chains}
\label{sec:SPTPhaseTransitions}

In this section, we consider various SO($N$)-symmetric spin chains that realize the field theory in Eq.~\eqref{eq:NIsingCFTHam} in the low-energy limit.
The numerical analysis of  these spin chains or spin ladder problems from $N=3$ to $N=6$ will give complementary results to the coupled-Ising chains simulations of Sec.~\ref{sec:IsingNum} on the nature of the phase transition in the field theory given by Eq.~\eqref{eq:NIsingCFTHam}.

\subsection{$N=3$: Spin-$1/2$ two-leg ladder with bond alternation}
\label{sec:BondAlternatingLadder}

As discussed in Ref.~\cite{YJWang00}, the $N=3$ case can be realized as the effective low-energy theory for a spin-$1/2$ two-leg ladder Hamiltonian with bond alternations,
\begin{align} \label{eq:DimerizedLadder}
& H_\mathrm{SO(3)} = \sum_{j=1}^L \bigl[ J_1 (\bm{S}_{j,1} \cdot \bm{S}_{j+1,1} + \bm{S}_{j,2} \cdot \bm{S}_{j+1,2}) \nonumber \\
&\quad + J_\perp \bm{S}_{j,1} \cdot \bm{S}_{j,2} 
+\delta_c (-1)^j (\bm{S}_{j,1} \cdot \bm{S}_{j+1,1} +\bm{S}_{j,2} \cdot \bm{S}_{j+1,2}) \nonumber \\
&\quad +\delta_s (-1)^j (-\bm{S}_{j,1} \cdot \bm{S}_{j+1,1} +\bm{S}_{j,2} \cdot \bm{S}_{j+1,2}) \bigr], 
\end{align}
where $\bm{S}_{j,a} = (S^x_{j,a}, S^y_{j,a}, S^z_{j,a})$ is the spin-$1/2$ operator acting on the $j$th site of the $a$th leg. 
Here, we suppose $J_1 > 0$.
Since each rung contains two spin-$1/2$'s on which SU($2$) symmetry only acts in integer-spin representation, we can conclude that the ladder model has an SO($3$) symmetry per rung.
When $J_\perp = \delta_c = \delta_s = 0$, the low-energy theory of the model is described by two copies of the SU(2)$_1$ WZW CFT or four copies of the Ising CFT with marginally irrelevant couplings \cite{Shelton96, Nersesyan97,Allen-S-97}.
In terms of the latter description, the rung exchange coupling is represented for $J_\perp \ll J_1$ as 
\begin{align}
\bm{S}_{j,1} \cdot \bm{S}_{j,2} \sim -\sum_{a=1}^3 i\xi^a_R \xi^a_L +3 i\xi^4_R \xi^4_L,
\end{align}
whereas the bond alternations are represented for $\delta_c, \delta_s \ll J_1$ as
\begin{align}
(-1)^j (\bm{S}_{j,1} \cdot \bm{S}_{j+1,1} +\bm{S}_{j,2} \cdot \bm{S}_{j+1,2}) &\sim \mu_1 \mu_2 \mu_3 \mu_4, \\
(-1)^j (\bm{S}_{j,1} \cdot \bm{S}_{j+1,1} -\bm{S}_{j,2} \cdot \bm{S}_{j+1,2}) &\sim \sigma_1 \sigma_2 \sigma_3 \sigma_4.
\end{align}
Here, $\xi^a_{R/L}$ are Majorana fermion fields and $\sigma_a$ ($\mu_a$) are order (disorder) parameter fields on the Ising CFT.

When $J_\perp >0 $, the fourth Majorana fermion $\xi^4_{R/L}$ has a negative mass gap and acquires a finite expectation value $\langle \sigma_4 \rangle \neq 0$. 
Integrating out $\xi^4_{R/L}$ yields Eq.~\eqref{eq:NIsingCFTHam} for $N=3$ as the effective low-energy theory with $m \sim J_\perp$ and $\lambda_1 \sim -\delta_s$.
Although there is generally a nonzero marginal coupling $\lambda_2$, determining its precise form in terms of microscopic parameters is technically involved.
Since the value of $\lambda_2$ does not qualitatively affect the nature of the phase transition, we neglect it for the remainder of this section.
The effective theory enjoys the SO($3$) symmetry inherited from the lattice model. 
As discussed in Sec.~\ref{sec:CoupledIsing} and predicted in Ref.~\cite{YJWang00}, we expect a continuous phase transition, which belongs to the universality class of the SU(2)$_1$ WZW CFT, between the ordered and disordered phases driven by the two relevant operators $\lambda_1$ and $m$. 
Indeed, this has been confirmed by several numerical studies \cite{Martin-Delgado98, Kotov99, Cabra99, Nakamura03}. 
Furthermore, the ordered and disordered phases actually correspond to a trivial phase $(J_\perp \gg |\delta_s|)$ and a nontrivial SPT phase $(J_\perp \ll |\delta_s|)$, respectively, which are protected by the SO($3$) symmetry and distinguished by the presence of spin-$1/2$ edge modes. 

When $J_\perp <0$, the Majorana fermion $\xi^4_{R/L}$ has a positive mass gap and acquires a finite expectation value for the disorder-parameter field $\langle \mu_4 \rangle \neq 0$. 
Even in this case, integrating out $\xi^4_{R/L}$ and applying the Kramers-Wannier duality transformation, $m \to -m$ and $\sigma_a \leftrightarrow \mu_a$, yields the same effective theory as in Eq.~\eqref{eq:NIsingCFTHam} for $N=3$ with 
$m \sim -J_\perp$ and $\lambda_1 \sim \delta_c$.
Now, the effective theory can be regarded as a spin-$1$ Heisenberg chain near the SO(3)$_1 \sim$ SU(2)$_2$ critical point and perturbed by a bond alternation.
This model also exhibits a phase transition between a trivial phase ($-J_\perp \ll |\delta_c|$) and 
a nontrivial SPT phase ($-J_\perp \gg |\delta_c|$), which are again protected by the SO($3$) symmetry and distinguished by spin-$1/2$ edge modes \cite{Orignac02}.
As expected, the phase transition belongs to the SU(2)$_1$ universality class \cite{Affleck87,Totsuka95a}, which has been confirmed by numerical studies (see, e.g., Refs.~\cite{Kato94, Yamamoto95a, Yamamoto95b, Totsuka95}).

We note that these topological phase transitions are most commonly understood by a semiclassical analysis based on the O($3$) nonlinear sigma model with a topological $\theta$ term.
Since the bond alternation continuously changes the topological angle $\theta$ from 0 to $2\pi$, a phase transition from a trivial to a nontrivial SPT phase is expected.
It has been predicted by Affleck and Haldane \cite{Affleck85, Affleck87} that this transition corresponds to $\theta=\pi$ and is described by the SU(2)$_1$ WZW CFT.
Thus, our analysis based on the $N=3$ coupled Ising chains is consistent with these results.

\subsection{$N=4$: Spin-1/2 two-leg ladder with an SU(2)$\times$ SU(2) symmetry}
\label{sec:SU2SU2Ladder}

For $N=4$, the Hamiltonian in Eq.~\eqref{eq:NIsingCFTHam} describes the low-energy effective theory for the following spin-$1/2$ two-leg ladder model
\begin{align} \label{eq:SO4LadderHam}
H_\mathrm{SO(4)} &= \sum_{j=1}^L \bigl[ J_1 (\bm{S}_{j,1} \cdot \bm{S}_{j+1,1} + \bm{S}_{j,2} \cdot \bm{S}_{j+1,2}) \nonumber \\
&\quad +J_4 (\bm{S}_{j,1} \cdot \bm{S}_{j+1,1}) (\bm{S}_{j,2} \cdot \bm{S}_{j+1,2}) \nonumber \\
&\quad +\delta_c (-1)^j (\bm{S}_{j,1} \cdot \bm{S}_{j+1,1} +\bm{S}_{j,2} \cdot \bm{S}_{j+1,2}) \nonumber \\
&\quad +\delta_s (-1)^j (-\bm{S}_{j,1} \cdot \bm{S}_{j+1,1} +\bm{S}_{j,2} \cdot \bm{S}_{j+1,2}) \bigr].
\end{align}
This Hamiltonian is similar to Eq.~\eqref{eq:DimerizedLadder} discussed above, but the rung exchange coupling $J_\perp$ is now replaced by a dimer-dimer interaction $J_4$ involving four spins.
In order to make a connection with the SO($4$) symmetry explicit, we introduce the generators of the SO($4$) Lie algebra, $L^{ab}_j = -L^{ba}_j$ $(1 \leq a < b \leq 4)$, by \cite{Tu2008}
\begin{align}
\begin{split}
& L^{12}_j = -S^z_{j,2} -S^z_{j,1},\quad 
L^{13}_j = S^x_{j,2} - S^x_{j,1}, \\
& L^{14}_j = -S^y_{j,2} - S^y_{j,1}, \quad
L^{23}_j = S^y_{j,2} - S^y_{j,1}, \\
& L^{24}_j = S^x_{j,2} + S^x_{j,1}, \quad
L^{34}_j = S^z_{j,2} - S^z_{j,1},
\end{split}
\end{align}
which satisfy
\begin{align} \label{eq:SO4algebra}
[L^{ab}_j, L^{cd}_k] = i\delta_{jk} (\delta_{ad} L^{bc}_k + \delta_{bc} L^{ad}_k - \delta_{ac} L^{bd}_k - \delta_{bd} L^{ac}_k).
\end{align}
The Hamiltonian in Eq.~\eqref{eq:SO4LadderHam} is rewritten as
\begin{align} \label{eq:SO4LadderHambis}
H_\mathrm{SO(4)} &= \sum_{j=1}^L \Biggl[ \left( \frac{J_1}{2} +\frac{J_4}{8} +\frac{\delta_c}{2}(-1)^j \right) \sum_{a<b} L^{ab}_j L^{ab}_{j+1} \nonumber \\
&\quad +\frac{J_4}{8} \left( \sum_{a<b} L^{ab}_j L^{ab}_{j+1} \right)^2 \nonumber \\
&\quad -\frac{\delta_s}{2} (-1)^j (L^{12}_j L^{34}_{j+1} - L^{13}_j L^{24}_{j+1} + L^{14}_j L^{23}_{j+1} \nonumber \\
&\quad +L^{34}_j L^{12}_{j+1} -L^{24}_j L^{13}_{j+1} + L^{14}_j L^{23}_{j+1}) \Biggr].
\end{align}
Here, the first two terms represent bilinear and biquadratic interactions between neighboring ``SO($4$) spins,'' which are obviously SO($4$)-invariant. The last term is allowed by the semisimple structure of SO(4) $\sim$ SU(2) $\times$ SU(2).

We suppose $J_1 >0$ in the following discussion. In the absence of the bond alternations $\delta_c=\delta_s=0$, the model is invariant under one-site translation along the chain, $\bm{S}_{j,a} \to \bm{S}_{j+1,a}$.
The $J_4=0$ point corresponds to two decoupled spin-$1/2$ Heisenberg chains with SO(4)$_1$ criticality ($c=2$) and also to the Reshetikhin point of the SO(4) model in Eq.~\eqref{eq:SO4LadderHambis}, which is integrable \cite{Reshetikhin83, Reshetikhin85, Tu2008}.
It has been known that the dimer-dimer interaction $J_4$ induces spontaneous dimerization with broken translation symmetry and a two-fold ground-state degeneracy \cite{Nersesyan97, Kolezhuk98}. 
The model realizes a columnar dimer phase for $J_4<0$ \cite{Hijii13, Robinson19} and a staggered dimer phase for $0<J_4<4J$ \cite{Pati98, Azaria99, Itoi00, Yamashita00}, in addition to a gapless critical phase described by the SU(4)$_1$ WZW CFT for $J_4 \geq 4J$ \cite{Pati98, Azaria99, Itoi00, Yamashita00}. 
The staggered dimer phase of model \eqref{eq:SO4LadderHambis} with $\delta_c=\delta_s=0$ has a special point that admits an exact MPS description with a two-fold ground state degeneracy \cite{Kolezhuk98,Tu2008}.
This phase enjoys a nonzero string order parameter \cite{Kolezhuk98,Tu2008},
\begin{align}
\langle O_\textrm{str} \rangle = 
\lim_{ | i-j | \rightarrow \infty} 
\left< L^{ab}_i \prod_{k=i+1}^{j-1} \exp(i \pi L^{ab}_{k}) L^{ab}_j \right> = \frac{1}{4}.
\end{align}
The non-vanishing of this string order parameter, together with the existence of edge states transforming 
in the spinor representations of the SO(4) group, has been interpreted as evidence for a doubly degenerate SO(4) SPT phase, dubbed SPT$^2$ phase in Ref.~\cite{Cai2025}.

The columnar bond alternation $\delta_c$ breaks the one-site translation symmetry to the two-site one $\bm{S}_{j,a} \to \bm{S}_{j+2,a}$ while maintains a reflection symmetry $\bm{S}_{j,1} \leftrightarrow \bm{S}_{j,2}$. 
It thus induces a phase transition from the staggered dimer phase with spontaneously broken reflection symmetry to the columnar dimer phase with no symmetry breaking. 
In contrast, the staggered bond alternation $\delta_s$ breaks the one-site translation symmetry to a glide symmetry $\bm{S}_{j,1(2)} \to \bm{S}_{j+1,2(1)}$, and induces a phase transition from the columnar dimer phase with spontaneously broken glide symmetry to the staggered dimer phase with no spontaneous symmetry breaking.

The ladder Hamiltonian in Eq.~\eqref{eq:SO4LadderHam} has the SU(2) $\times$ SU(2) $\sim$ SO(4) symmetry related to spin rotations on each chain, which is required by the CFT Hamiltonian in Eq.~\eqref{eq:NIsingCFTHam}.
As discussed in Sec.~\ref{sec:BondAlternatingLadder}, the model for $J_4 = \delta_c = \delta_s = 0$ is described by four copies of the Ising CFT with marginally irrelevant couplings in the low-energy limit.
Since the dimer-dimer interaction is represented for $J_4 \ll J_1$ as \cite{Nersesyan97}
\begin{align}
(\bm{S}_{j,1} \cdot \bm{S}_{j+1,1}) (\bm{S}_{j,2} \cdot \bm{S}_{j+1,2}) \sim -\sum_{a=1}^4 i\xi^a_R \xi^a_L,
\end{align}
the low-energy effective theory of Eq.~\eqref{eq:SO4LadderHam} for $\delta_c=0$ precisely matches the coupled Ising CFT Hamiltonian in Eq.~\eqref{eq:NIsingCFTHam} for $N=4$ with coupling constants $m \sim J_4$ and $\lambda_1 \sim -\delta_s$. The ordered and disordered phases in Eq.~\eqref{eq:NIsingCFTHam} thus correspond to the columnar dimer phase for $J_4 \gg |\delta_s|$ and the (symmetric) staggered dimer phase for $-J_4 \ll |\delta_s|$, respectively, with negative $J_4$.
When $\delta_s = 0$, we can also obtain the same effective theory with $m \sim -J_4$, $\lambda_1 \sim \delta_c$, and $\lambda_2 \sim \gamma$ after the Kramers-Wannier duality transformation.
The ordered and disordered phases then correspond to the staggered dimer phase for $J_4 \gg |\delta_c|$ and the (symmetric) columnar dimer phase for $J_4 \ll |\delta_c|$, respectively, with positive $J_4$.
Thus, the phase transitions from the dimer phases with spontaneous symmetry breaking to the distinct dimer phases with no symmetry breaking are described by the coupled Ising CFT Hamiltonian in Eq.~\eqref{eq:NIsingCFTHam} for $N=4$, which we expect to have a first-order transition.

We here focus on the regime of negative $J_4$ and set $J_1=1$, $J_4=-1$, and $\delta_c=0$. 
While $J_4$ is not perturbativelly small and it is subtle to justify Eq.~\eqref{eq:NIsingCFTHam} as an effective low-energy theory, we make this choice of parameters to clearly see the signature of the first-order transition; 
since critical SO($N$) spin chains inevitably have a marginal coupling $\lambda_2$, which is positive and sizably large, we need enormous bond dimensions (or long length scales) to access the true asymptotic behavior caused by the flow to a negatively infinite marginal coupling when we are close to the critical point, as suggested by the RG analysis in Sec.~\ref{sec:RG} (see also the related discussion in Appendix~\ref{app:Additional}).
By varying the staggered bond alternation $\delta_s$, we numerically study the phase transition from the columnar dimer phase with spontaneously broken glide symmetry to the staggered dimer phase with no symmetry breaking.

We employ the VUMPS method and exploit the SU(2) $\times$ SU(2) symmetry of the model. 
We impose the two-site periodicity for the infinite MPS (i.e., each unit cell contains four spin-$1/2$'s) and keep bond dimensions up to $\chi=12800$.
Conservation of the SU(2) $\times$ SU(2) symmetry requires that the virtual space of an MPS must be in representations of SU(2) $\times$ SU(2) and these representations for the virtual spaces must be consistent under fusion with the representations for the physical space.
In the current model, the physical space is in the ($1/2$,$1/2$) representation and the left-to-right virtual spaces must be in either (H,H)-to-(I,I) or (H,I)-to-(I,H) representations, up to one-site translation, where H and I refer to the half-integer and integer representations, respectively.
Which representation is legitimate for the variational MPS is determined by which representation minimizes the ground-state energy density for the MPS.
We found that the energy density for $\chi=12800$ is minimized when we choose the (H,H)-to-(I,I) representation for $\delta_s \leq 0.192$ and the (H,I)-to-(I,H) representation for $\delta_s \geq 0.194$.
We note that this switching of the representation does not necessarily imply the presence of a first-order transition, since physical quantities computed with different representations of the MPS can smoothly change for a sufficiently large bond dimension when the transition is of continuous-type.
As shown in Appendix~\ref{app:son}, the signatures of the first-order transition have been found even if we fix the representation of the virtual spaces.

\begin{figure}[tb]
\includegraphics[width=0.48\textwidth]{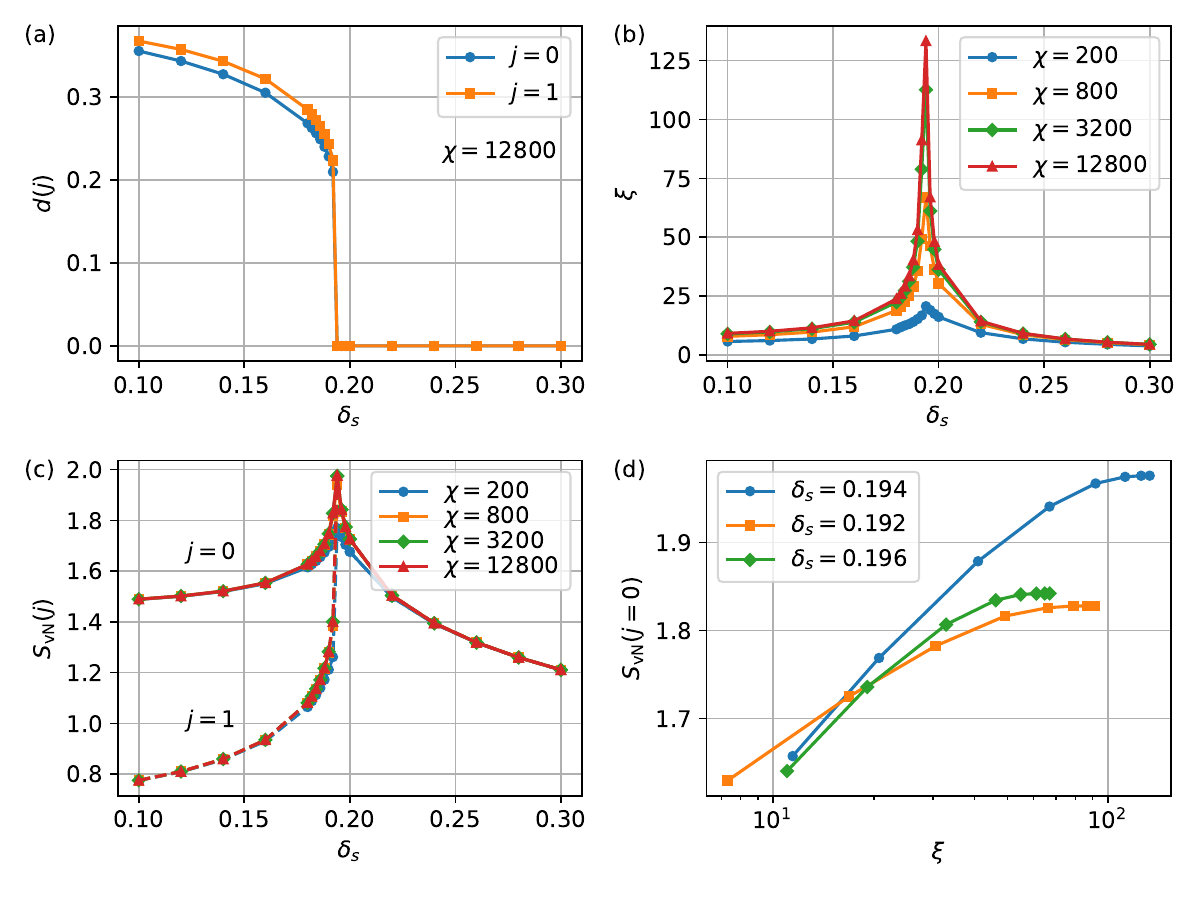}
\caption{Numerical results for the SO(4)-symmetric spin ladder in Eq.~\eqref{eq:SO4LadderHam} with $J_1=1$, $J_4=-1$, and $\delta_c=0$.
(a) Dimerization order parameters $d(j)$, (b) correlation length $\xi$, (c) von Neumann entanglement entropy $S_\textrm{vN}(j)$ are plotted as functions of $\delta_s$. 
(d) $S_\textrm{vN}(j=0)$ are plotted against the correlation length $\xi$ at the vicinity of the transition.}
\label{fig:SO4Ladder}
\end{figure}

In Fig.~\ref{fig:SO4Ladder}(a), we show dimerization order parameters,
\begin{align} \label{eq:DimerOrderParam}
d(j) \equiv |\langle \bm{S}_{j-1,1} \cdot \bm{S}_{j,1} - \bm{S}_{j,2} \cdot \bm{S}_{j+1,2} \rangle|,
\end{align}
which become finite in the columnar dimer phase for $\delta_s \lesssim 0.194$ whereas they vanish in the staggered dimer phase for $\delta_s \gtrsim 0.194$.
It is notable that the change in the dimerization order parameters is very abrupt at the transition $\delta_s \sim 0.194$, signaling its first-order nature. 
In Figs.~\ref{fig:SO4Ladder}(b) and (c), we show the correlation length $\xi$ and the von Neumann entanglement entropy $S_\textrm{vN}(j)$ at the cuts $j=0,1$ for several bond dimensions, both of which peak at the transition point $\delta_s \sim 0.194$.
In order to see a possible critical behavior at the transition, we study the scaling form of the von Neumann entanglement entropy $S_\textrm{vN}(j=0)$ with the correlation length $\xi$, which is given by $S_\textrm{vN} \sim (c/6)\ln \xi$ with central charge $c$ at a conformal invariant critical point.
As shown in Fig.~\ref{fig:SO4Ladder}(d), the entanglement entropy saturates to a constant value even right at the transition. 
This gives a further support on the first-order transition in the field theory given by Eq.~\eqref{eq:NIsingCFTHam} for $N=4$, consistent with the numerical results in Sec.~\ref{sec:IsingNum}.

\subsection{$N=5$: SO(5) bilinear-biquadratic spin chain with bond alternation}

For $N=5$, one can consider a generic bilinear-biquadratic SO(5) model, for which the phase diagram has been studied previously~\cite{Tu2008, Alet2011},
\begin{align}\label{eq:so5}
H'_\mathrm{SO(5)} = \cos\theta \sum_{j=1}^{L} \sum_{a<b} L_j^{ab} L_{j+1}^{ab} + \sin\theta \sum_{j=1}^{L} \left(\sum_{a<b} L_j^{ab} L_{j+1}^{ab} \right)^2,
\end{align}
where $L^{ab}$ ($1\leq a < b \leq 5$) are the 10 generators of the five-dimensional vector representation of SO(5), which satisfy the SO(5) Lie algebra commutation relations as in Eq.~\eqref{eq:SO4algebra}. 
The model in Eq.~\eqref{eq:so5} enjoys a rich phase diagram, which includes a nontrivial SO(5) SPT phase, when $\theta_R < \theta < \pi/4$ [$\theta_R \equiv \arctan ((N-4)/(N-2)^2) = \arctan(1/9)$], with edge states that belong to the spinor representation of SO(5) with dimension 4. 
The Reshetikhin point at $\theta = \theta_R$ corresponds to the SO(5)$_1$ critical point and separates the SO(5) SPT phase from a two-fold degenerate dimerized phase, which contains the purely bilinear model with $\theta=0$ \cite{Tu2008, Alet2011}.

A field theory in terms of five Majorana fermions $\xi^a_{R/L} (a= 1,\ldots, 5)$ has been introduced in Ref.~\cite{Alet2011} to describe these phases and the Reshetikhin point.
The starting point there was the low-energy approach for the half-filled spin-$3/2$ Hubbard model, which has an exact SO(5) symmetry \cite{Wu2003}. 
As discussed in Refs.~\cite{Alet2011, Hao_Orus2011}, the vicinity of the Reshetikhin point can be described by the SO(5) massive Majorana field theory
\begin{align} \label{eq:SO5CFTHam}
H'_\mathrm{SO(5)} &\sim -\frac{iv}{4\pi} \int dx \sum_{a=1}^5 (\xi_R^a \partial_x \xi_R^a -\xi_L^a \partial_x \xi^a_L) \nonumber \\
&\quad -im \int dx \sum_{a=1}^5 \xi_R^a \xi_L^a,
\end{align}
where we have neglected a marginal four-fermion term.
The Reshetikhin point corresponds to $m=0$ with critical properties in the SO(5)$_1$ universality class with central charge $c=5/2$.
In the dimerized phase, the Majorana fermions have a positive mass $m>0$ and the SO(5) dimerization operator 
\begin{align}
D_j = (-1)^j \sum_{a<b} L_j^{ab} L_{j+1}^{ab}  \sim \prod_{a=1}^{5} \mu_a
\end{align}
condenses; $\langle D_j \rangle \ne 0$, which signals a two-fold degenerate ground state with spontaneous breaking of the one-site translation symmetry.
In contrast, the phase with $m < 0$ (i.e., the ordered phase of the underlying Ising model) is a nondegenerate gapped phase, which is a nontrivial SO(5) SPT phase, that is, an SO(5) generalization of the Haldane phase. The stable edge state of that phase, which belongs to the spinorial representation of the SO(5) group, can be described by means of this Majorana formulation \cite{Alet2011}.

Starting from this SPT phase and adding an explicit dimerization to Eq.~\eqref{eq:so5}, one can drive a transition to a trivial nondegenerate gapped phase for a large dimerization.
Specifically, we consider the following model:
\begin{align}\label{eq:so5dimer}
H_\mathrm{SO(5)} &=& \cos\theta \sum_{j=1}^{L} (1+(-1)^j\delta)\sum_{a<b} L_j^{ab} L_{j+1}^{ab} \phantom{space}\nonumber \\
&+& \sin\theta \sum_{j=1}^{L} (1+(-1)^j\delta)\left(\sum_{a<b} L_j^{ab} L_{j+1}^{ab} \right)^2,
\end{align}
i.e., the same Hamiltonian as in Eq.~\eqref{eq:so5} except that even (respectively odd) bonds have an amplitude multiplied by $(1+\delta)$ (respectively $(1-\delta)$).

Using Eq.~\eqref{eq:SO5CFTHam} and the correspondence $D_i  \sim \prod_{a=1}^{5} \mu_a$, we find that the transition is governed by the field theory
\begin{align} \label{eq:SO5CFTHamper}
H_\mathrm{SO(5)} &\sim -\frac{iv}{4\pi} \int dx \sum_{a=1}^5 (\xi_R^a \partial_x \xi_R^a -\xi_L^a \partial_x \xi^a_L) \nonumber \\
&\quad -im \int dx \sum_{a=1}^5 \xi_R^a \xi_L^a  + \delta \int dx \prod_{a=1}^{5} \mu_a,
\end{align}
with $m<0$.
The two strongly relevant perturbations have an antagonistic effect as in Eq.~\eqref{eq:NIsingCFTHam} with $N=5$, since the mass term favors an ordered phase ($m<0$), i.e., the SO(5) SPT phase, whereas the explicit dimerization operator with coupling constant $\delta$ stabilizes a disordered phase with $\langle \mu_a \rangle \ne 0$. 
In fact, the model in Eq.~\eqref{eq:SO5CFTHamper} can be directly mapped onto Eq.~\eqref{eq:NIsingCFTHam} with $N=5$ by performing a Kramers-Wannier duality transformation.

We numerically investigate the model in Eq.~\eqref{eq:so5dimer} with explicit dimerization by using an exact mapping onto a spin-$2$ model (see Ref.~\cite{Alet2011}), so that SU(2) symmetry can be imposed in our iDMRG simulations.
The SO(5) SPT phase can be characterized by nontrivial edge states, spin-$3/2$ edge states in this formulation, which is reflected in iDMRG simulations as having a lower entanglement entropy when using appropriate virtual spaces.
In practice, as in Sec. \ref{sec:SU2SU2Ladder}, one performs simulations for different classes of virtual spaces (e.g., half-integer vs integer SU(2) representations) and check if the infinite MPS is injective or not.
This allows us to compute properly the correlation length $\xi$ using an appropriate scaling~\cite{Rams2018}.
Technical details on similar computations can be found, e.g., in Ref.~\cite{Capponi2025}.

\begin{figure}[!htb]
\includegraphics[width=0.49\columnwidth]{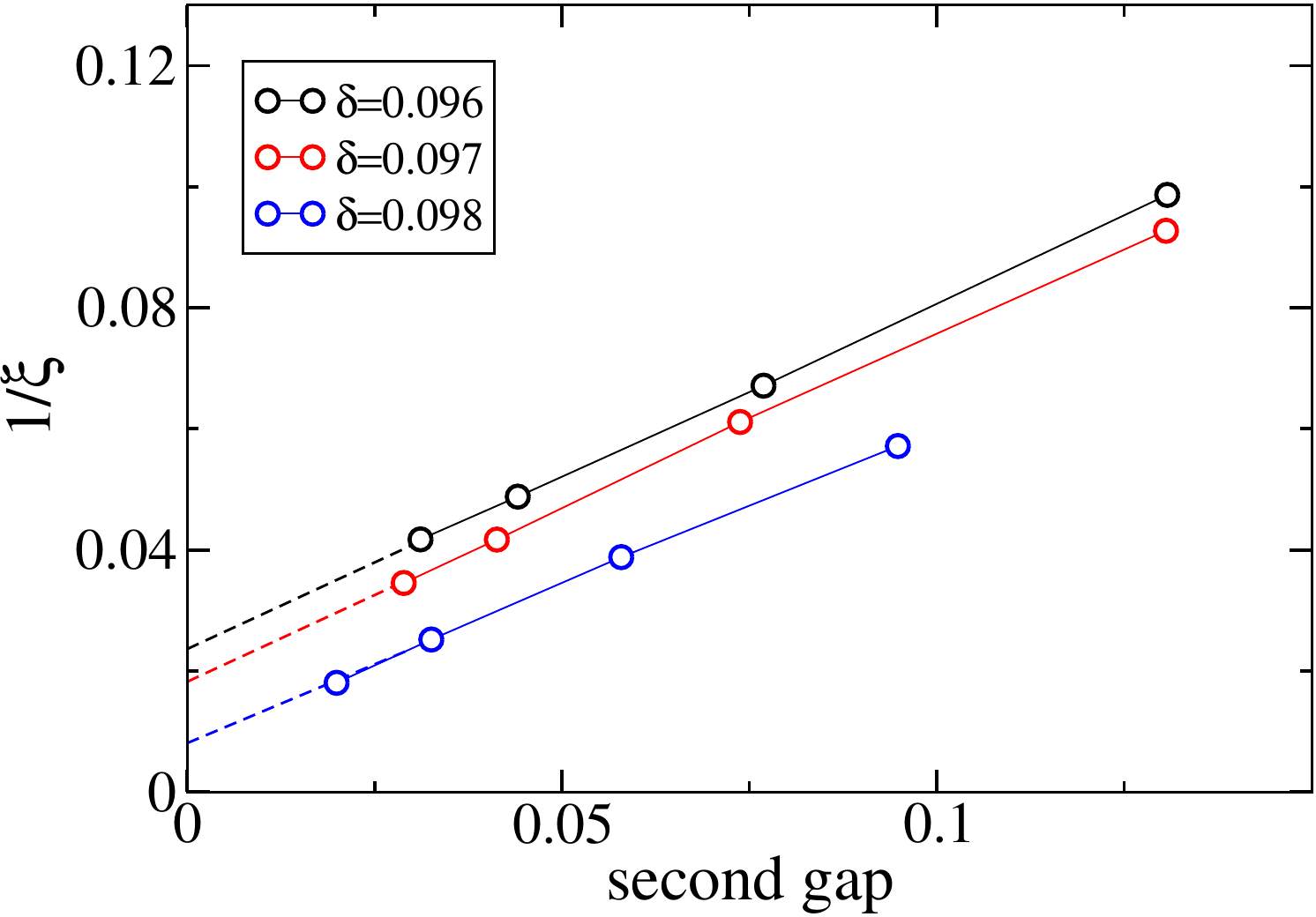}
\includegraphics[width=0.49\columnwidth]{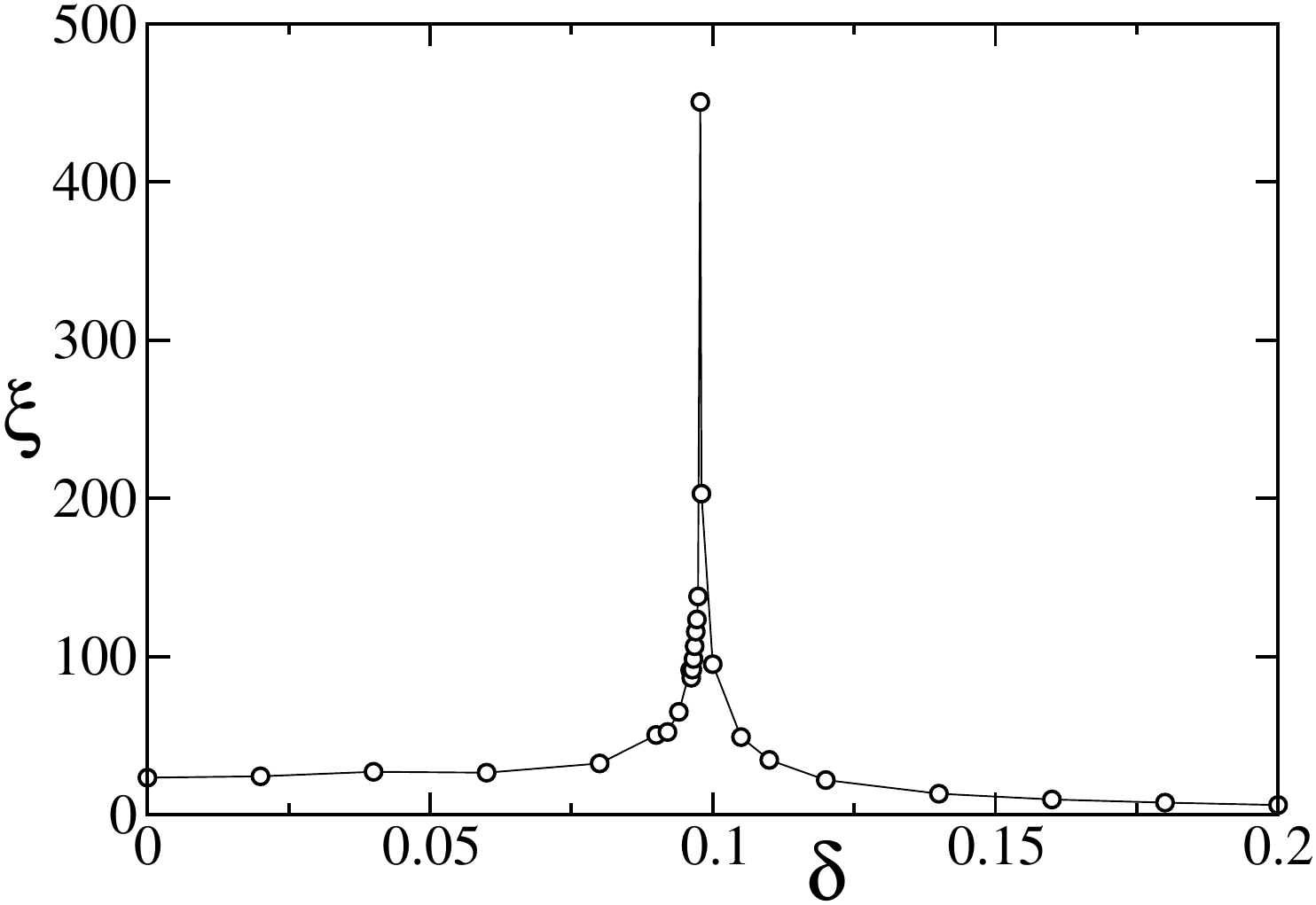}
\caption{Numerical results for the SO(5)-symmetric spin chain in Eq.~\eqref{eq:so5dimer} with $\theta=8^\circ$ using data with a discarded weight down to $10^{-5}$.
(left) Extrapolation of the inverse of the correlation length vs the second gap for several values of $\delta$ around the transition.
(right) Extrapolated correlation length $\xi$ vs dimerization $\delta$.}
\label{fig:xi_SO5}
\end{figure}

In order to start close to the critical point, we have fixed $\theta = 8^\circ$ and used a minimal unit cell of two sites.
For $\delta=0$ (uniform case), we recover the SO(5) SPT phase with four-fold degenerate edge states corresponding to $S=3/2$ in the SU(2) language and a finite correlation length.
For such a case, we use half-integer virtual spaces (H) on the left edge of the unit cell.
When $\delta \geq 0.098$, we need to switch to integer ones (I) in order to minimize the entanglement entropy.
By performing the appropriate extrapolation using the second gap of the transfer matrix spectrum~\cite{Rams2018} and using the correct virtual space sector [see Fig.~\ref{fig:xi_SO5}(a)], one can obtain the behavior of the correlation length vs dimerization, as shown in Fig.~\ref{fig:xi_SO5}(b).
While we again observe a very large value of $\xi$ at the transition, our extrapolated results remain finite.
Overall, this indicates that the field theory in Eq.~\eqref{eq:NIsingCFTHam} for $N=5$ also has a first-order transition \footnote{
We have additionally checked that the extrapolated correlation length remains finite at first-order transitions for a tricritical Ising chain proposed in Ref.~\cite{OBrien2018} by performing a similar numerical analysis (data not shown).
}.

\subsection{$N=6$: SO(6) bilinear-biquadratic spin chain with bond alternation}

We now turn to the $N=6$ case and consider a related lattice model as in $N=5$. 
Since SO(6) and SU(4) have the same Lie algebra, we take advantage of a natural formulation in terms of an SU(4) spin chain where each site transforms as the six-dimensional self-conjugate fully antisymmetric irreducible representation, which corresponds to the vector representation of SO(6).
In this respect, we consider a bilinear-biquadratic model with explicit bond alternation, 
\begin{align}
 H_\mathrm{SO(6)} 
& =   - J_p \sum_{i=1}^{L} \left\{ \cos \theta (S^{A}_{2i-1}  S^{A}_{2i})  + \frac{\sin\theta}{4} (S^{A}_{2i-1}  S^{A}_{2i}) ^2 \right\} \nonumber \\
& \phantom{ = } 
+ \sum_{i=1}^{L} \left\{ \cos \theta (S^{A}_{2i}  S^{A}_{2i+1}) + \frac{\sin\theta}{4} (S^{A}_{2i}  S^{A}_{2i+1}) ^2\right\}  \; ,
\label{eq:so6}
\end{align}
where $S^{A}_{i}$ ($A=1, \ldots, 15$) denote the SU($4$) spin operators in the six-dimensional fully-antisymmetric self-conjugate representation ($\bm{6}$).
These operators, which are the generators of the SU($4$) algebra in this representation, are normalized as $\mathrm{Tr}(S_{i}^{A} S_{j}^{B}) = 2 \delta_{ij} \delta^{AB}$.
Throughout this subsection, summation over the index $A$ is implicitly assumed.
The uniform case ($J_p= -1$) has been studied in Ref.~\cite{Affleck1991} by means of various approaches.
The model exhibits a critical point when $\theta = \theta_R = \arctan{1/2}$, which corresponds to the Reshetikhin point ~\cite{Reshetikhin85} with SO(6)$_1$ critical behavior and central charge $c=3$.
This critical point separates a spontaneously dimerized phase ($-3\pi/4 < \theta < \theta_R$) and a nontrivial C-breaking phase ($\theta_R < \theta < \pi/4$), both of which are two-fold degenerate \cite{Affleck1991}.
The C-breaking phase is characterized by a nonzero three-site order parameter \cite{Affleck1991},
\begin{align}
C_j  =  {\rm tr} \left(S_{j}  S_{j+1} S_{j+2} \right) + (j \leftrightarrow j+2),
\end{align}
where $(S_j)_{\alpha \beta} = S^A_j (T^A)_{\alpha \beta}$.
Here, $T^A$ are the SU($4$) generators in the fundamental representation, which are normalized as $\mathrm{tr}(T^A T^B) = \delta^{AB}/2$, and the trace $\mathrm{tr}(\cdot)$ is taken over the fundamental SU($4$) indices $\alpha,\beta=1,\ldots,4$. 
This phase breaks the ``charge-conjugation'' (C) symmetry, under which the spin operators transforms as $S_j \to -S_j^T$.
The C-order parameter stabilizes a staggered charge-conjugation order that can be understood from an exact MPS description for $\theta = \arctan{(2/3)}$ where $\langle C_i \rangle = (-1)^i 10/3 $ \cite{Affleck1991}.
This phase can be interpreted as an SO(6) SPT phase with edge states that belong to the fundamental ($\bm{4}$) and anti-fundamental ($\bar{\bm{4}}$) representations of the SU(4) group \cite{Capponi2026}. 

The effective field theory that governs the physical properties of the model in Eq.~\eqref{eq:so6} near the Reshetikhin point
with SO(6)$_1$ criticality is very similar to the SO(5) case in Eq.~\eqref{eq:SO5CFTHamper}, apart from the fact that we need here six Majorana fermions: 
\begin{align} \label{eq:SO6CFTHamper}
H_\mathrm{SO(6)} &= -\frac{iv}{4\pi} \int dx \sum_{a=1}^6 (\xi_R^a \partial_x \xi_R^a -\xi_L^a \partial_x \xi^a_L) \nonumber \\
&\quad -im \int dx \sum_{a=1}^6 \xi_R^a \xi_L^a  + g \int dx \prod_{a=1}^{6} \mu_a,
\end{align}
with $g \sim J_p +1$. 
In the absence of $g$, the phase for $m < 0$ corresponds to the C-breaking phase characterized by a finite expectation value of $C_j  \sim \prod_{a=1}^{6} \sigma_a$, whereas the phase for $m > 0$ corresponds to a dimerized phase characterized by a dimerization order parameter $D_j  = (-1)^{j} S^{A}_{j} S^{A}_{j+1} \sim \prod_{a=1}^{6} \mu_a$ \cite{Hao_Orus2011}.
As in the SO(5) case,  the two strongly relevant perturbations favor different types of ordering: the mass term stabilizes the C-breaking phase, whereas the perturbation $g$ with scaling dimension $3/4$ leads to a trivial nondegenerate gapped singlet phase.
The nature of the resulting phase transition can be studied numerically by investigating the lattice model in Eq.~\eqref{eq:so6} near the Reshetikhin point.

\begin{figure}[!htb]
\includegraphics[width=\columnwidth]{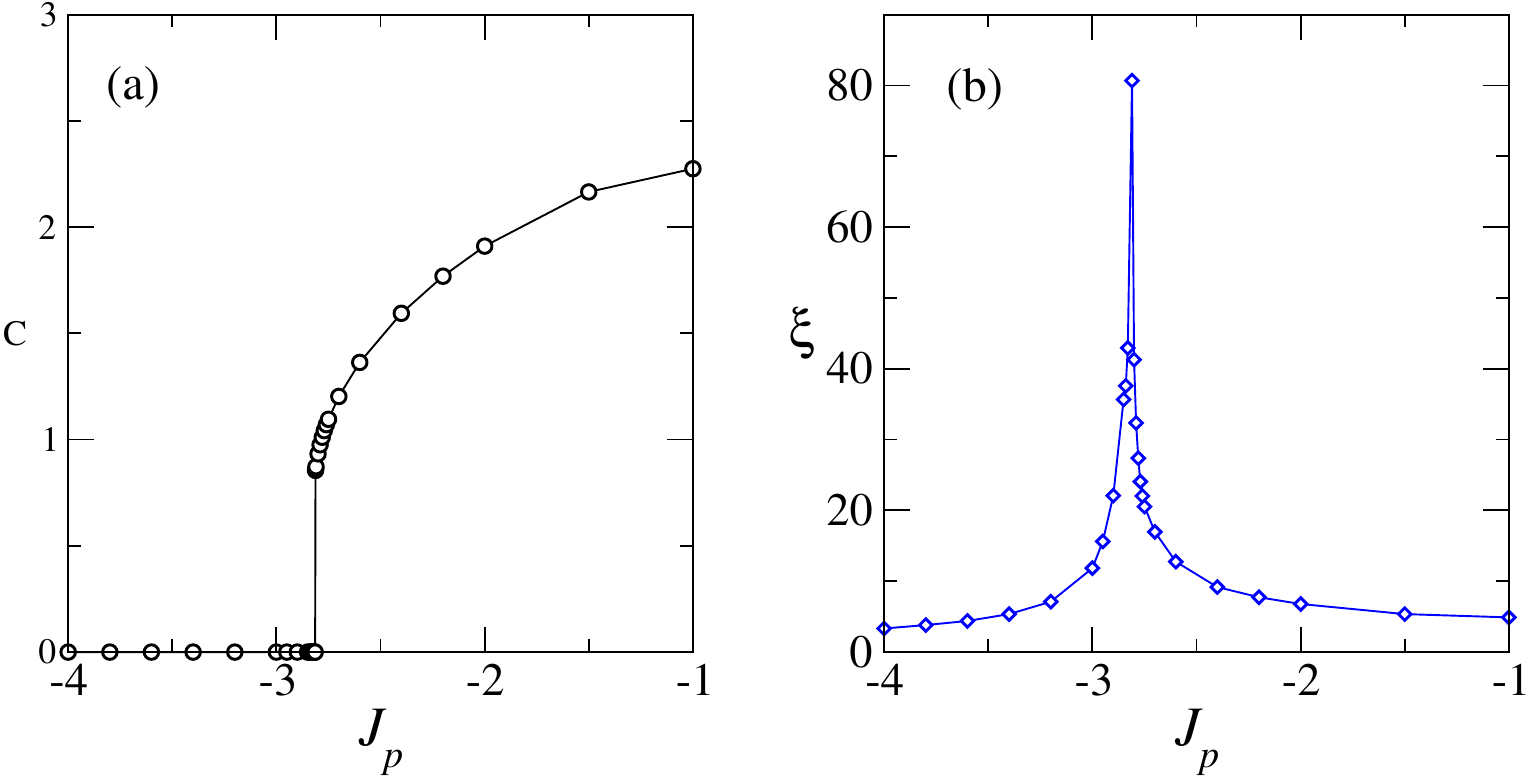}
\caption{Numerical results for the SO(6)-symmetric spin chain in Eq.~\eqref{eq:so6} with $\theta=0.175\pi$ using data with a discarded weight down to $10^{-5}$ and appropriate virtual spaces (see the main text). 
(a) C-breaking order parameter $C=|\langle C_i \rangle|$ vs $J_p$ . 
(b) Extrapolated correlation length $\xi$ vs $J_p$.}
\label{fig:xi_SO6}
\end{figure}

The full phase diagram of model \eqref{eq:so6} is currently under investigation~\cite{Capponi2026}, but we report here some results regarding the phase transition from the C-breaking phase to the trivial phase obtained when $J_p \neq -1$. We consider here $J_p<0$ and since the model is invariant by $J_p \leftrightarrow 1/J_p$, we restrict ourselves to $J_p<-1$.
We fix $\theta=0.175\pi$ ($\theta_R< \theta$) so that the system is in the C-breaking phase for $J_p=-1$.
This is confirmed both from getting a lower entanglement entropy using the appropriate virtual space at the left edge, i.e., $\bf{4}$ or $\bf{\bar{4}}$ [in contrast to other SU(4) irreducible representations having $0$ or $2 \pmod{4}$ boxes] and by measuring the C-breaking order parameter, which takes $\langle C_i \rangle \approx 1.14 (-1)^i $ at $J_p=-1$.
This changes suddenly when $J_p < -2.812$ where the adequate virtual spaces contain only irreducible representations with number of boxes multiple of 4 and $\langle C_i \rangle = 0$.
As shown in Fig.~\ref{fig:xi_SO6}, the order parameter jump is compatible with first-order transition, so that the correlation length $\xi$, obtained from an extrapolation similar to the ones performed previously, remains finite (albeit quite large) at the transition.
These results suggest that the field theory in Eq.~\eqref{eq:NIsingCFTHam} for $N=6$ also has a first-order transition.

\subsection{$N=6$: Spin-$1$ two-leg ladder with SO(3) $\times$ SO(3) symmetry}

We here consider a spin-$1$ two-leg  ladder described by the following Hamiltonian
\begin{align} \label{eq:Spin1BLBQLadder}
H &= \sum_{j=1}^L \bigl[ J_1 h_\theta(\bm{S}_{j,1} \cdot \bm{S}_{j+1,1}) +J_1 h_\theta(\bm{S}_{j,2} \cdot \bm{S}_{j+1,2}) \nonumber \\
&\quad +J_4 h_\theta(\bm{S}_{j,1} \cdot \bm{S}_{j+1,1}) h_\theta(\bm{S}_{j,2} \cdot \bm{S}_{j+1,2}) \bigr],
\end{align}
where $\bm{S}_{j,a}$ is the spin-1 operator acting on the $j$th site of the $a$th leg and $h_\theta(\bm{S}_i \cdot \bm{S}_j)$ represents a bilinear-biquadratic interaction parametrized by the angle $\theta \in (-\pi,\pi]$, 
\begin{align}
h_\theta(\bm{S}_i \cdot \bm{S}_j) = \cos \theta (\bm{S}_i \cdot \bm{S}_j) + \sin \theta (\bm{S}_i \cdot \bm{S}_j)^2.
\end{align}
When $\theta=-\pi/4$ and $J_4=0$, the model describes two decoupled critical spin-$1$ chains, known as the Takhtajan-Babujian chain, whose low-energy theory is given by the SU(2)$_2$ or SO(3)$_1$ WZW CFT with $c=3/2$.
This decoupled point is thus described by six copies of the critical Ising CFTs.
Slight deviations from this decoupled point are captured by the coupled Ising CFTs in Eq.~\eqref{eq:NIsingCFTHam} for $N=6$ with $m \propto \delta \theta \equiv \delta -\pi/4$ and $\lambda_1 \propto J_4$, although this model does not have the full SO(6) symmetry as in Eq. \eqref{eq:so6} but only its subgroup SO(3) $\times$ SO(3) and a $\mathbb{Z}_2$ symmetry interchanging two chains.
This anisotropy only shows up at the level of marginal interactions and will not modify the nature of the phase transition governed by the relevant perturbations $m$ and $\lambda_1$.
When a positive $m$ flows to the strong-coupling limit, the model can be smoothly deformed into two decoupled Haldane spin chains.  The corresponding ground state thus belongs to an SPT phase protected by the SO(3) $\times$ SO(3) symmetry. On the other hand, a strong four-spin interaction $J_4$ can induce a columnar (staggered) dimer phase with spontaneous breaking of the translation symmetry for $J_4<0$ ($J_4>0$). 

Here, we focus on the negative $J_4$ regime and set $J=1$ and $\theta=-0.1\pi$ such that the model is in the doubled Haldane phase in the decoupling limit $J_4=0$. 
We then numerically study the phase transition to the columnar dimer phase by the VUMPS method with fully utilizing the SU(2) $\times$ SU(2) symmetry.
Similarly to the SU(2) $\times$ SU(2) symmetric spin ladder analyzed in Sec.~\ref{sec:SU2SU2Ladder}, we impose the two-site periodicity for the infinite MPS and keep bond dimensions up to $\chi=12800$.
In this model, the physical space is in the (1,1) representation of SU(2) $\times$ SU(2) and the virtual spaces must be in one of the (H,H), (I,I), or (H,I) representations, up to the interchange of the two chains.
We found that the ground-state energy density is minimized by the (I,I) representation for $J_4 \leq -0.382$ whereas by the (H,H) representation for $J_4 \geq -0.380$.

\begin{figure}[!tb]
\includegraphics[width=0.48\textwidth]{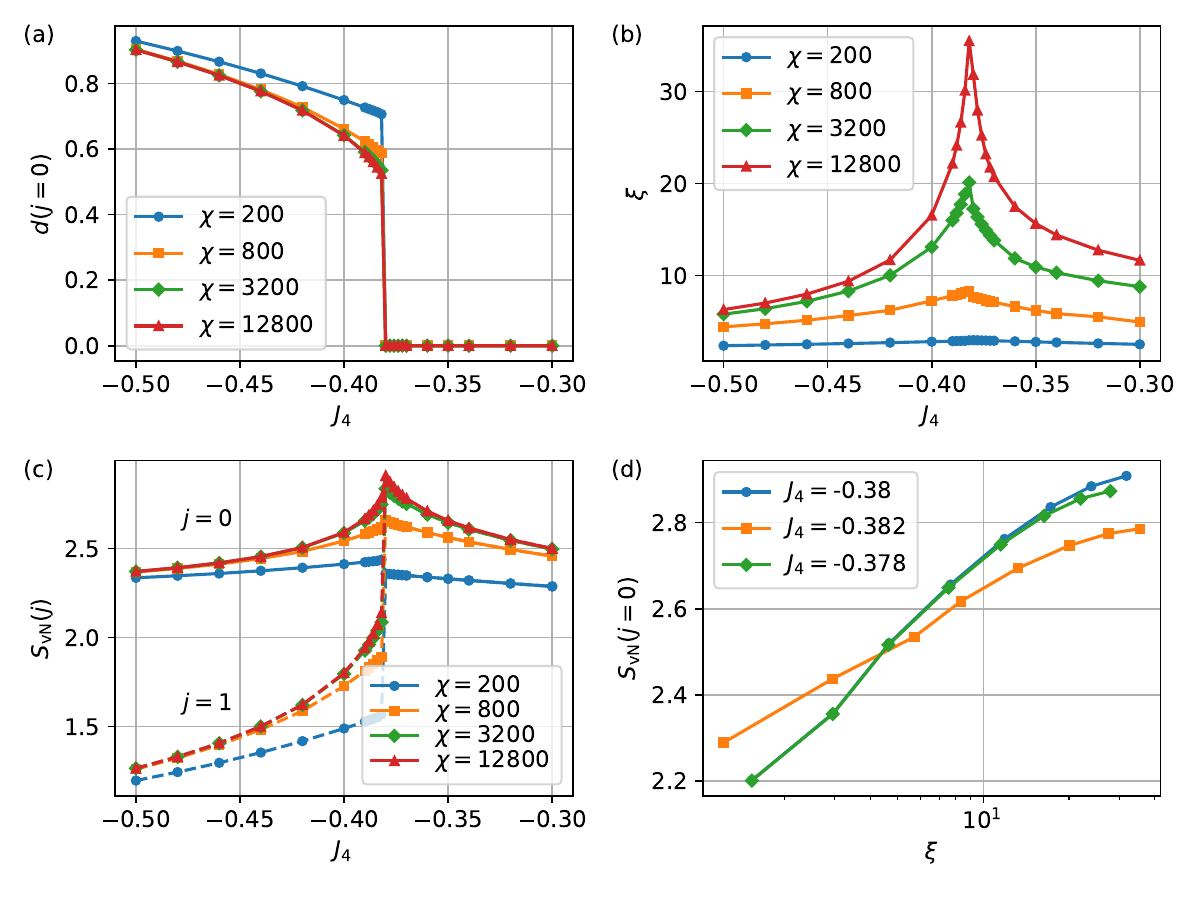}
\caption{Numerical results for the SO(3) $\times$ SO(3) symmetric ladder in Eq.~\eqref{eq:Spin1BLBQLadder} with $J_1=1$ and $\theta=-0.1\pi$.
(a) Dimerization order parameters $d(j=0)$, (b) correlation length $\xi$, and (c) von Neumann entanglement entropy $S_\textrm{vN}(j)$ are plotted as functions of $J_4$.
(d) $S_\textrm{vN}(j=0)$ are plotted against the correlation length $\xi$ at the vicinity of the transition.}
\label{fig:BLBQLadder}
\end{figure}

In Fig.~\ref{fig:BLBQLadder}(a), we show the dimerization order parameter defined in Eq.~\eqref{eq:DimerOrderParam}, which suddenly jumps to take a finite value around $J_4 \sim -0.38$, signaling the first-order transition. 
As shown in Figs.~\ref{fig:BLBQLadder}(b) and (c), both correlation length $\xi$ and entanglement entropy $S_\textrm{vN}(j)$ show a peak at $J_4 \sim -0.38$; in particular, the latter is accompanied by an abrupt increase for the lower-entangled bond $(j=1)$ across the transition.
As shown in Fig.~\ref{fig:BLBQLadder}(d), the entanglement entropy as a function of the correlation length appears to saturate to a constant value in the vicinity of $J_4 \sim 0.38$, signaling the gap opening across the transition. 
These results give another support that the effective theory in Eq.~\eqref{eq:NIsingCFTHam} for $N=6$ has a first-order transition.

\section{Conclusion}
\label{sec:Conclusion}

In this manuscript, we studied the nature of the phase transition driven by two competing relevant operators in the $N$-coupled Ising CFT [Eq.~\eqref{eq:NIsingCFTHamintro}]. 
This effective field theory governs many quantum phase transitions in the context of spin ladders 
and SO($N$)-symmetric spin chains.
Combining the perturbative RG analysis and numerical simulations using MPS, we provide compelling evidence that the transition is of the continuous type for $N=2$ and $N=3$ with Ising and four-state Potts universality, respectively, whereas of the first-order type for $4 \leq N < 16$.
We then discuss the implications of these results in the context of SO($N$)-symmetric spin chains, for which we numerically confirmed the presence of first-order transitions between SPT and dimerized phases for $4 \leq N \leq 6$.

An immediate application of our results is a refinement of an interesting conjecture by Verresen, Moessner, and Pollmann about the nature of the phase transitions between SPT phases \cite{Verresen17}.
They argue that if there exists a direct continuous phase transition between an SPT phase with $d$-dimensional edge states and a trivial phase, it should have a central charge $c \geq \log_2 d$.
For $N = 2n+1$, our field theory in Eq.~\eqref{eq:NIsingCFTHamintro} describes a direct transition between an SO($2n+1$) SPT phase with $2^{n}$-dimensional edge states and a trivial phase.
For $N=3$, their conjecture holds, as the transition belongs to the SU(2)$_1$ universality class with central charge $c=1$.
However, our results indicate that their assumption of the existence of a continuous phase transition does not hold for $N \geq 5$; the transition generally becomes first order in this case.
For $N=2n$ and $N \geq 4$, the field theory describes a direct transition between an SO($2n$) SPT phase with $2^{n-1}$-dimensional edge states and an extra $\mathbb{Z}_2$ symmetry breaking and a trivial phase which should also be first order.

While our numerical simulations have been performed on lattice models whose low-energy effective theory is expected to match the field theory in Eq.~\eqref{eq:NIsingCFTHamintro}, this correspondence is not well justified beyond a perturbative regime around the decoupled critical Ising models or SO($N$)$_1$ critical point.
Instead, one could more directly investigate the non-perturbative RG flow of the field theory \eqref{eq:NIsingCFTHamintro} by means of truncated conformal space approach \cite{Yurov-Z-90,James-K-L-R-T-18}.
It is also interesting to explore the nature of a nontrivial fixed point in the space of complex parameters \cite{Gorbenko18a, Gorbenko18b}, as suggested by our perturbative RG analysis. 
Such complex CFT might be realized by adding non-Hermitian perturbations to the first-order transitions of our lattice models for $N \geq 4$.

A possible extension is to consider the phase transitions in $N$ copies of the $q$-state Potts CFTs, which reduces to our model for $q=2$. 
This model bears renewed recent interest for $3 \leq q \leq 4$ and $N \geq 3$ and when the copies are coupled via thermal operators, as it is a candidate for a nontrivial irrational CFT \cite{Dotsenko99, Kousvos24} (see also Refs.~\cite{Antunes23, Antunes25a, Antunes25b}).
It is then natural to ask if the introduction of additional competing relevant perturbations leads to nontrivial fixed points described by CFT and how the nature of the transitions depends on $N$.

\acknowledgments

We would like to thank Keisuke Totsuka for fruitful discussions and collaboration on a related project.
We thank Yin-Chen He, Hong-Hao Tu, and Atsushi Ueda for valuable discussions.
The authors are supported by the IRP project ``Exotic Quantum Matter in Multicomponent Systems (EXQMS)'' from CNRS. 
Y.F. is supported by JSPS KAKENHI Grant No. JP24K06897.
The Flatiron Institute is a division of the Simons Foundation (L.D.). 
All MPS simulations are performed using TensorKit \cite{TensorKit} and MPSKit \cite{MPSKit} packages.
We thank CALMIP (grant 2025-P0677) and GENCI (project A0170500225) for computer resources.

\section*{Data availability}

The data used in the figures of this manuscript are openly available \cite{data}.

\appendix

\section{Details of RG equations}

\subsection{Derivation of RG equations} \label{app:DerivationRGEq}

We here consider the Euclidean action corresponding to the Hamiltonian in Eq.~\eqref{eq:NIsingCFTHam}, 
\begin{align}
S &= \int \frac{d\tau dx}{4\pi} \sum_{a=1}^N \left[ \xi^a_R (\partial_\tau -iv\partial_x) \xi^a_R +\xi^a_L (\partial_\tau +iv\partial_x) \xi^a_L \right] \nonumber \\
&\quad +\frac{v}{2\pi} \sum_{j=0,1,2} G_j \int \frac{d\tau dx}{\alpha^{2-\Delta_j}} O_j(x),
\end{align}
where $\alpha$ is a short-distance cutoff at the lattice scale and $\Delta_j$ is the scaling dimension for the operator $O_j(x)$. 
The dimensionless coupling constants $G_j$ and scaling operators $O_j(x)$ are given by
\begin{align}
& G_0 = \frac{2\pi \alpha}{v} m, \quad G_1 = \frac{2\pi \alpha^{2-N/8}}{v} \lambda_1, \quad G_2 = \frac{2\pi}{v} \lambda_2, \\
& O_0 = \sum_{a=1}^N \varepsilon_a, \quad O_1 = \prod_{a=1}^N \sigma_a, \quad O_2 = \sum_{a<b} \varepsilon_a \varepsilon_b,
\end{align}
where we have introduced the thermal operator for the Ising CFT, 
\begin{align}
\varepsilon_a(x) \equiv -i\xi^a_R(x) \xi^a_L(x).
\end{align}
Using the operator-product expansion (OPE), we can write down the perturbative RG equations at one loop \cite{Cardy}, 
\begin{align}
\frac{dG_k}{dl} = (2-\Delta_k) G_k -\sum_{i,j} \frac{C_{ijk}}{2} G_i G_j.
\end{align}
In order to compute the OPE coefficients $C_{ijk}$, we use the OPEs for the primary operators in Ising CFT \cite{CFT},
\begin{subequations}
\begin{align}
\varepsilon(z,\bar{z}) \varepsilon(w,\bar{w}) & \sim \frac{1}{|z-w|^2}, \\
\varepsilon(z,\bar{z}) \sigma(w,\bar{w}) &\sim \frac{1}{2|z-w|} \sigma(w,\bar{w}), \\
\sigma(z,\bar{z}) \sigma(w,\bar{w}) &\sim \frac{1}{|z-w|^{1/4}} +\frac{1}{2} |z-w|^{3/4} \varepsilon(w,\bar{w}),
\end{align}
\end{subequations}
where we have introduced the complex coordinates $z=v\tau+ix$ and $\bar{z}=v\tau-ix$.
We then find the OPEs for the operators $O_j$, with an abuse of notation $O_j(z) \equiv O_j(z,\bar{z})$,
\begin{subequations}
\begin{align}
O_0(z) O_0(w) &\sim \frac{N}{|z-w|^2} +2O_2(w), \\
O_0(z) O_1(w) &\sim \frac{N}{2|z-w|} O_1(w), \\
O_0(z) O_2(w) &\sim \frac{N-1}{|z-w|^2} O_0(w), \\
O_1(z) O_1(w) &\sim \frac{1}{|z-w|^{N/4}} +\frac{1}{2|z-w|^{N/4-1}} O_0(w) \nonumber \\
&\quad +\frac{1}{4|z-w|^{N/4-2}} O_2(w), \\
O_1(z) O_2(w) &\sim \frac{N(N-1)}{8|z-w|^2} O_1(w), \\
O_2(z) O_2(w) &\sim \frac{N(N-1)}{2|z-w|^4} +\frac{2(N-2)}{|z-w|^2} O_2(w),
\end{align}
\end{subequations}
which yield the RG equations given in Eq.~\eqref{eq:RGEq}.

\subsection{Nontrivial fixed points of RG equations} \label{app:FixedPointRGEq}

Substituting $N=16-\epsilon$ into the RG equations in Eq.~\eqref{eq:RGEq}, we find six nontrivial fixed points $(G_0^*, G_1^*, G_2^*)$. 
In the leading order of $\epsilon$, they are given by
\begin{subequations}
\begin{align}
\textrm{FP}_1^\pm: & \left( \pm i \frac{\sqrt{14}}{15} +O(\epsilon), \, 0, \, \frac{1}{15} +O(\epsilon) \right), \\
\textrm{FP}_2^\pm: & \left( O(\epsilon^2), \, \pm i \frac{\sqrt{7} \epsilon}{60} +O(\epsilon^2), \, \frac{\epsilon}{240} +O(\epsilon^2) \right), \\
\textrm{FP}_3^\pm: & \left( -\frac{225}{1798} +O(\epsilon), \, \pm i \frac{\sqrt{898}}{899} +O(\epsilon), \, \frac{30}{899} +O(\epsilon) \right).
\end{align}
\end{subequations}
Thus, while $\textrm{FP}_2^\pm$ are perturbatively accessible, none of them lie in the real parameter space of the coupling constants $G_j$. 

\onecolumngrid
\section{Additional numerical results}
\label{app:Additional}

In this appendix, we provide several additional details for our numerical results presented in the main text.

\subsection{Ashkin-Teller model}
\label{app:AT}

\begin{figure}[tb]
\includegraphics[width=0.48\textwidth]{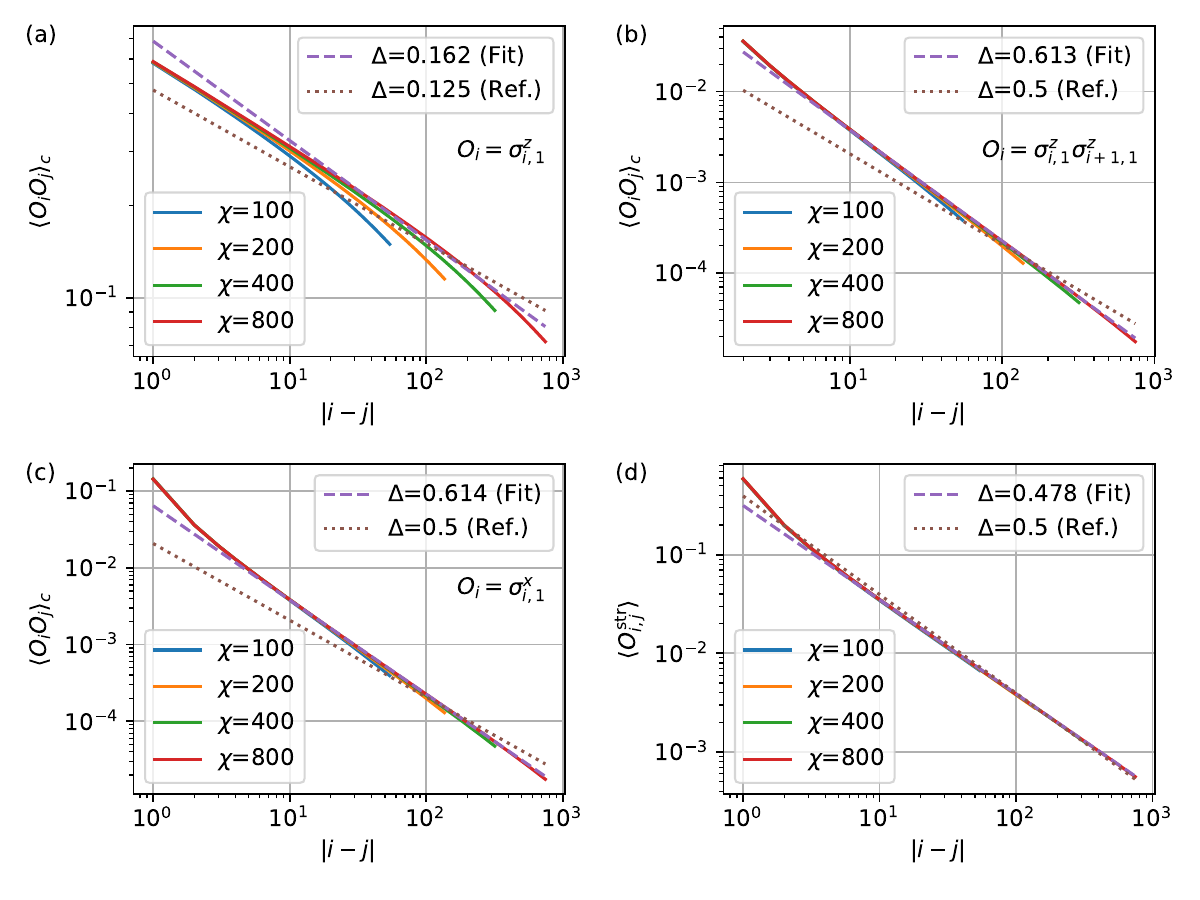}
\caption{
Numerical results for the Ashkin-Teller model at the four-state Potts criticality in Eq.~\eqref{eq:AT}.
Two-point connected correlation functions $\langle O_i O_j \rangle_c$ for (a) $O_i = \sigma^z_{i,1}$, (b) $O_i = \sigma^z_{i,1} \sigma^z_{i+1,1}$, and (c) $O_i = \sigma^x_{i,1}$. 
The string correlation function $\langle O^\textrm{str}_{i,j} \rangle$ is also shown in (d).
The blue dashed lines are fitting functions of the power-law form $C |i-j|^{-2\Delta}$, whereas the orange ones are those with exact exponents of the four-state Potts CFT shown as references.
\label{fig:AshkinTellerCC}
}
\end{figure}

In Sec.~\ref{sec:IsingNum}, we claimed that the coupled Ising Hamiltonian in Eq.~\eqref{eq:NIsingHam} for $N=3$ has a continuous phase transition described by the four-state Potts CFT.
Although the scaling analysis for the entanglement entropy confirms the expected central charge $c=1$, those for the correlation functions of local operators strongly deviate from the asymptotic behaviors expected from the CFT.
We here perform a similar numerical analysis for the Ashkin-Teller model at the four-state Potts criticality,
\begin{align} \label{eq:AT}
H_\textrm{AT} &= -\sum_{j=1}^L \left( \sigma^z_{j,1} \sigma^z_{j+1,1} +\sigma^x_{j,1} +\sigma^z_{j,2} \sigma^z_{j+1,2} +\sigma^x_{j,2} \right. \nonumber \\
&\quad \left. +\sigma^z_{j,1} \sigma^z_{j+1,1} \sigma^z_{j,2} \sigma^z_{j+1,2} +\sigma^x_{j,1} \sigma^x_{j,2} \right),
\end{align}
by using the VUMPS algorithm and by implementing the $\mathbb{Z}_2 \times \mathbb{Z}_2$ symmetry generated by $\prod_{j=1}^L \sigma^x_{j,1}$ and $\prod_{j=1}^L \sigma^x_{j,2}$.
As shown in Fig.~\ref{fig:AshkinTellerCC}, we find the critical exponents $\Delta \sim 0.16$ for $O_i = \sigma^z_{i,1}$ and $\Delta \sim 0.6$ for $O_i = \sigma^z_{i,1} \sigma^z_{i+1,1}$ and $\sigma^x_{i,1}$.
As shown in Table~\ref{tab:Coupled3Ising}, these results quantitatively match the numerical results for the $N=3$ coupled Ising chains but strongly deviate from the exact values $\Delta_\sigma=1/8$ and $\Delta_\epsilon=1/2$ expected for the four-state Potts CFT.
This indicates that even the original Ashkin-Teller model, which is exactly solvable, bears the strong deviation from the true asymptotic behaviors for finite bond dimensions, which is due the logarithmic correction by a marginally irrelevant operator \cite{Alcaraz87, Cardy86}.
On the other hand, as shown in Fig.~\ref{fig:AshkinTellerCC}(d), the string correlation function,
\begin{align}
\langle O^\textrm{str}_{i,j} \rangle = \left< \sigma^z_{i,1} \left( \prod_{k=i+1}^{j-1} \sigma^x_{k,2} \right) \sigma^z_{j,1} \right>,
\end{align}
receives a less severe correction and gives the exponent $\Delta \sim 0.478$, which is reasonably close to the exact value $\Delta_g=1/2$.
These results indicate that the $N=3$ coupled Ising Hamiltonian in Eq.~\eqref{eq:NIsingHam} at the phase transition shares quantitatively similar scaling properties with the Ashkin-Teller model at four-state Potts criticality and suggest that they are both in the same universality class.

\subsection{Coupled Ising chains}
\label{app:Ising}

In Sec.~\ref{sec:IsingNum}, we presented numerical results for the coupled Ising Hamiltonian in Eq.~\eqref{eq:NIsingHam} for $K=0$ and $g=0.5$.
We here provide additional results for the other choices of the parameters: $(K,g) = (0,2)$, $(\pm 0.3, 0.5)$, and $(\pm 0.3, 2)$.

\begin{figure*}[tb]
\includegraphics[width=\textwidth]{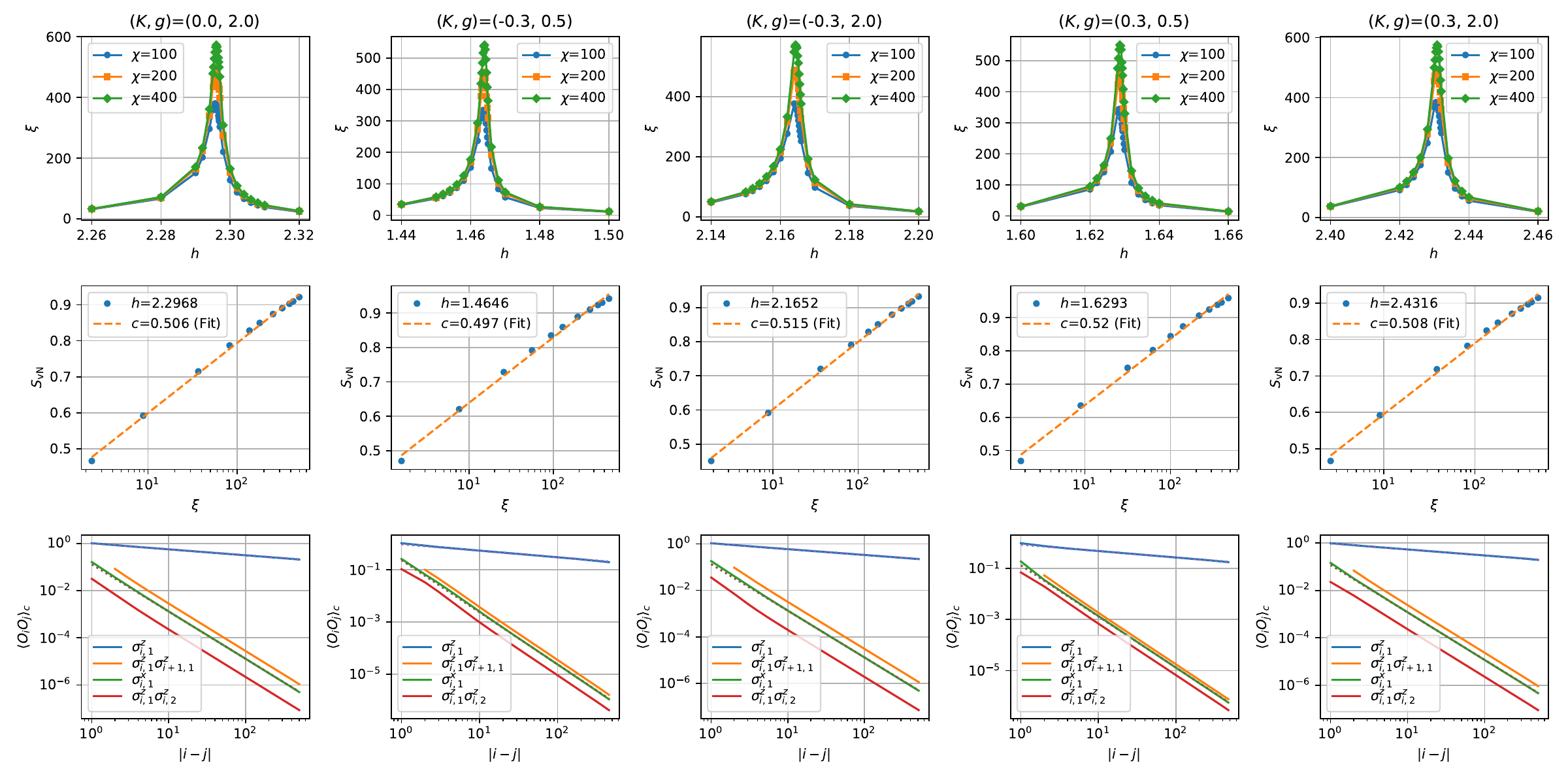}
\caption{
Numerical results for the $N=2$ coupled Ising chains in Eq.~\eqref{eq:NIsingHam} with $(K,g) = (0,2)$, $(-0.3,0.5)$, $(-0.3,2)$, $(0.3,0.5)$, and $(0.3,2)$. 
Top panels show the correlation length $\xi$ as functions of $h$. 
Middle panels show the von Neumann entanglement entropy $S_\textrm{vN}$ plotted against
$\xi$ near the transition points and fitting functions of the form $c/6 \log \xi + c_0$ (dashed lines).
Bottom panels show the connected correlation functions $\langle O_i O_j \rangle_c$ for $O_i
= \sigma^z_{i,1}$, $\sigma^z_{i,1} \sigma^z_{i+1,1}$, $\sigma^x_{i,1}$, and $\sigma^z_{i,1}
\sigma^z_{i,2}$ at the transition points and reference curves of the form $C |i-j|^{-2\Delta}$ with $\Delta=0.125$ (blue dashed lines) and $\Delta=1$ (orange dashed lines).
\label{fig:Coupled2IsingSuppl}
}
\end{figure*}

In Fig.~\ref{fig:Coupled2IsingSuppl}, we show the correlation length $\xi$, von Neuman entanglement entropy $S_\textrm{vN}$, and connected correlation functions $\langle O_i O_j \rangle_c$ for $O_i = \sigma^z_{i,1}$, $\sigma^z_{i,1} \sigma^z_{i+1,1}$, $\sigma^x_{i,1}$, and $\sigma^z_{i,1} \sigma^z_{i,2}$, all of which are computed for the $N=2$ coupled Ising Hamiltonian \eqref{eq:NIsingHam} using the iDMRG algorithm with $\mathbb{Z}_2 \times \mathbb{Z}_2^C$ symmetry.
For all choices of the parameter sets, we find a phase transition signaled by the sharp peak of $\xi$, at which we can extract the central charge $c \sim 0.5$ and the critical exponents $\Delta \sim 0.125$ for the order-parameter field and $\Delta \sim 1$ for the thermal field, as summarized in Table~\ref{tab:Coupled2Ising}.
These results are consistent with a continuous phase transition described by the Ising CFT.

\begin{figure*}[tb]
\includegraphics[width=\textwidth]{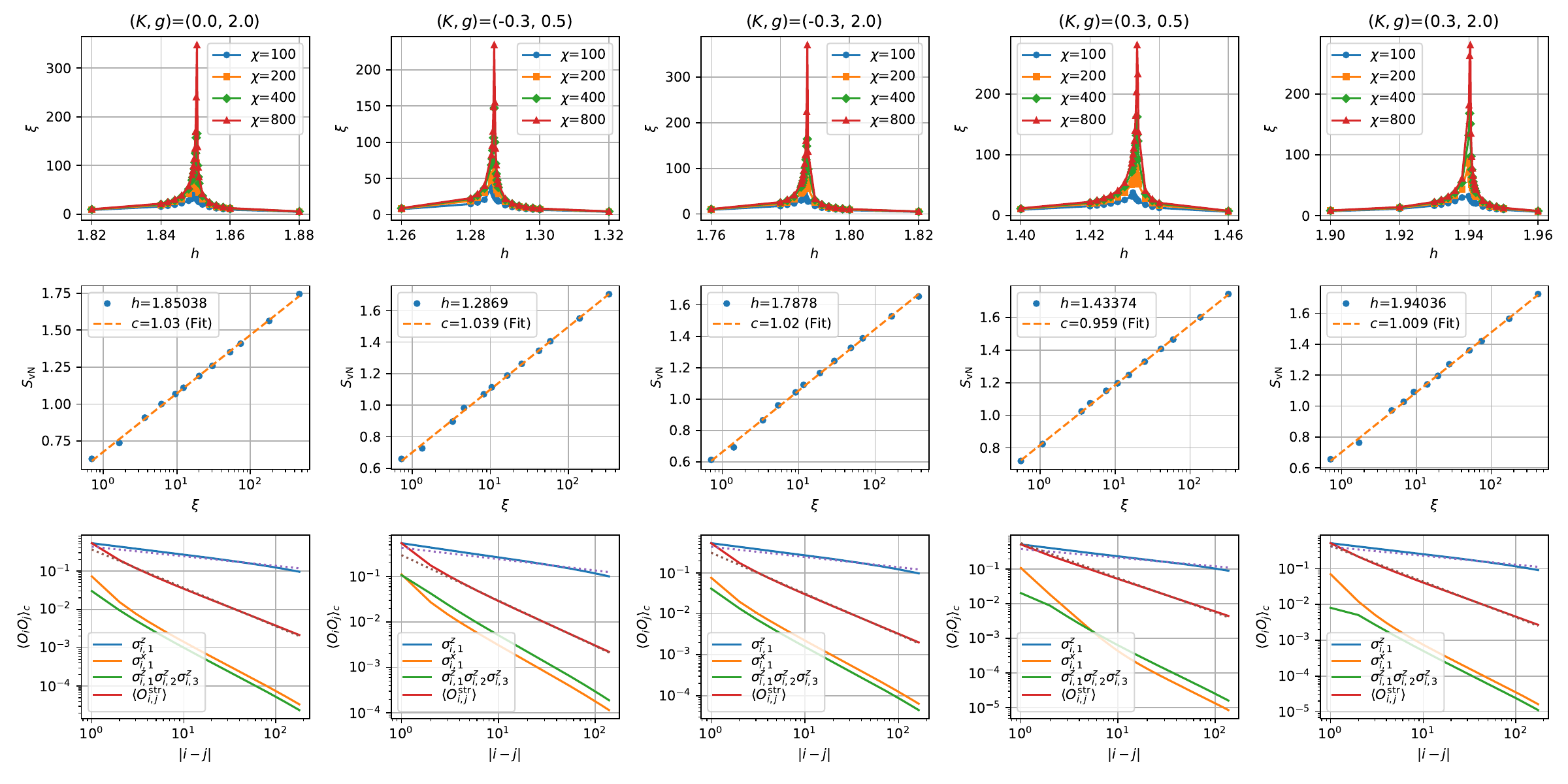}
\caption{
Numerical results for the $N=3$ coupled Ising chains in Eq.~\eqref{eq:NIsingHam} with $(K,g) = (0,2)$, $(-0.3,0.5)$, $(-0.3,2)$, $(0.3,0.5)$, and $(0.3,2)$. 
Top panels show the correlation length $\xi$ as functions of $h$. 
Middle panels show the von Neumann entanglement entropy $S_\textrm{vN}$ plotted against
$\xi$ near the transition points and fitting functions of the form $c/6 \log \xi + c_0$ (dashed lines).
Bottom panels show the connected correlation functions $\langle O_i O_j \rangle_c$ for $O_i
= \sigma^z_{i,1}$, $\sigma^x_{i,1}$, and $\sigma^z_{i,1} \sigma^z_{i,2} \sigma^z_{i,3}$ and
the string correlation function $\langle O^\textrm{str}_{i,j} \rangle$ at the transition
points and reference curves of the form $C |i-j|^{-2\Delta}$ with $\Delta=0.125$ (blue dashed lines) and $\Delta=0.5$ (orange dashed lines).
\label{fig:Coupled3IsingSuppl}
}
\end{figure*}

In Fig.~\ref{fig:Coupled3IsingSuppl}, we show the results for the $N=3$ coupled Ising Hamiltonian in Eq.~\eqref{eq:NIsingHam} using the VUMPS algorithm with $\mathbb{Z}_2^2$ symmetry.
Similarly to the $N=2$ case, for all choices of the parameter sets, a phase transition is signaled by the sharp peak of $\xi$.
While the central charge extracted from the entanglement entropy at the transition point is close to $c \sim 1$, the connected correlation functions show rather strong deviation from the expected behavior, as seen from the critical exponents summarized in Table~\ref{tab:Coupled3Ising}.
However, since the different sets of parameters give quantitatively similar results, which are also comparable with the Ashkin-Teller model at four-state Potts criticality as discussed above, we conclude that the $N=3$ model has a continuous phase transition described by the four-state Potts CFT.

\begin{figure}[tb]
\includegraphics[width=0.48\textwidth]{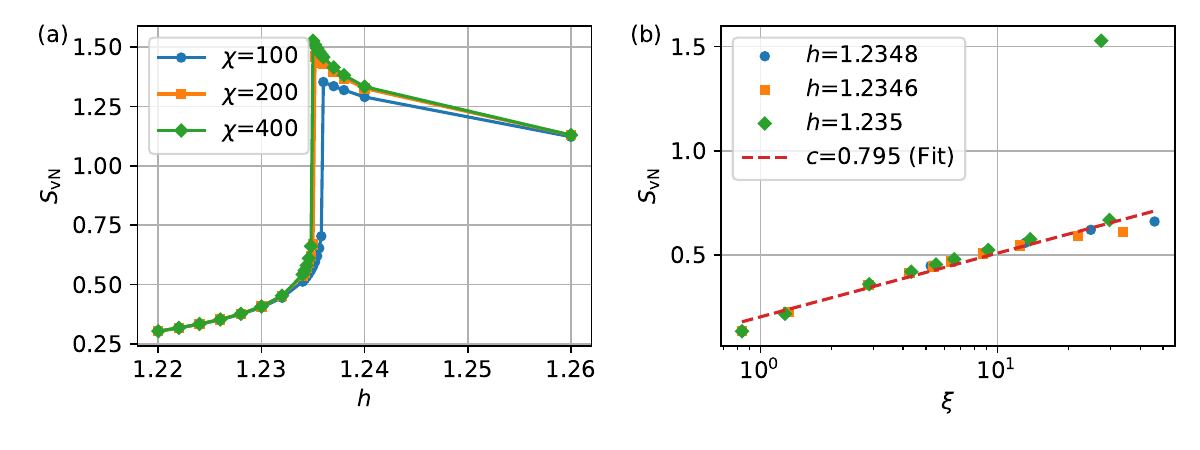}
\caption{
Numerical results for the $N=4$ coupled Ising chains in Eq.~\eqref{eq:NIsingHam} with $(K,g) = (0,0.5)$ using non-symmetric MPS. 
The von Neumann entanglement entropy is plotted against (a) the magnetic field $h$ and (b) the correlation length $\xi$. 
The dashed line is a fitting function of the form $(c/6) \log \xi + c_0$ (dashed lines).
\label{fig:Coupled4IsingSupplNosym}
}
\end{figure}

For $N=4$, we first show the VUMPS results for \emph{non-symmetric} MPS in Fig.~\ref{fig:Coupled4IsingSupplNosym}.
Compared with the result obtained by MPS with explicit $\mathbb{Z}_2^3$ symmetry (see Fig.~\ref{fig:Coupled4Ising}), the ordered phase for $h < h_c$ has a lower entanglement entropy because of the collapse of the cat state.
In this case, right at or \emph{below} the transition $h_c \leq 1.2348$ (namely, in the ordered phase), the entanglement entropy shows a tendency to saturate to a constant value for large correlation lengths. 
On the other hand, slightly inside the \emph{disordered} phase, the entanglement entropy shows a sudden jump as the bond dimension is increased.
These results suggest a first-order transition accompanied by a level crossing, as similarly observed for the symmetric MPS, albeit the reversed $h$-dependence of the entanglement entropy.

\begin{figure*}[tb]
\includegraphics[width=\textwidth]{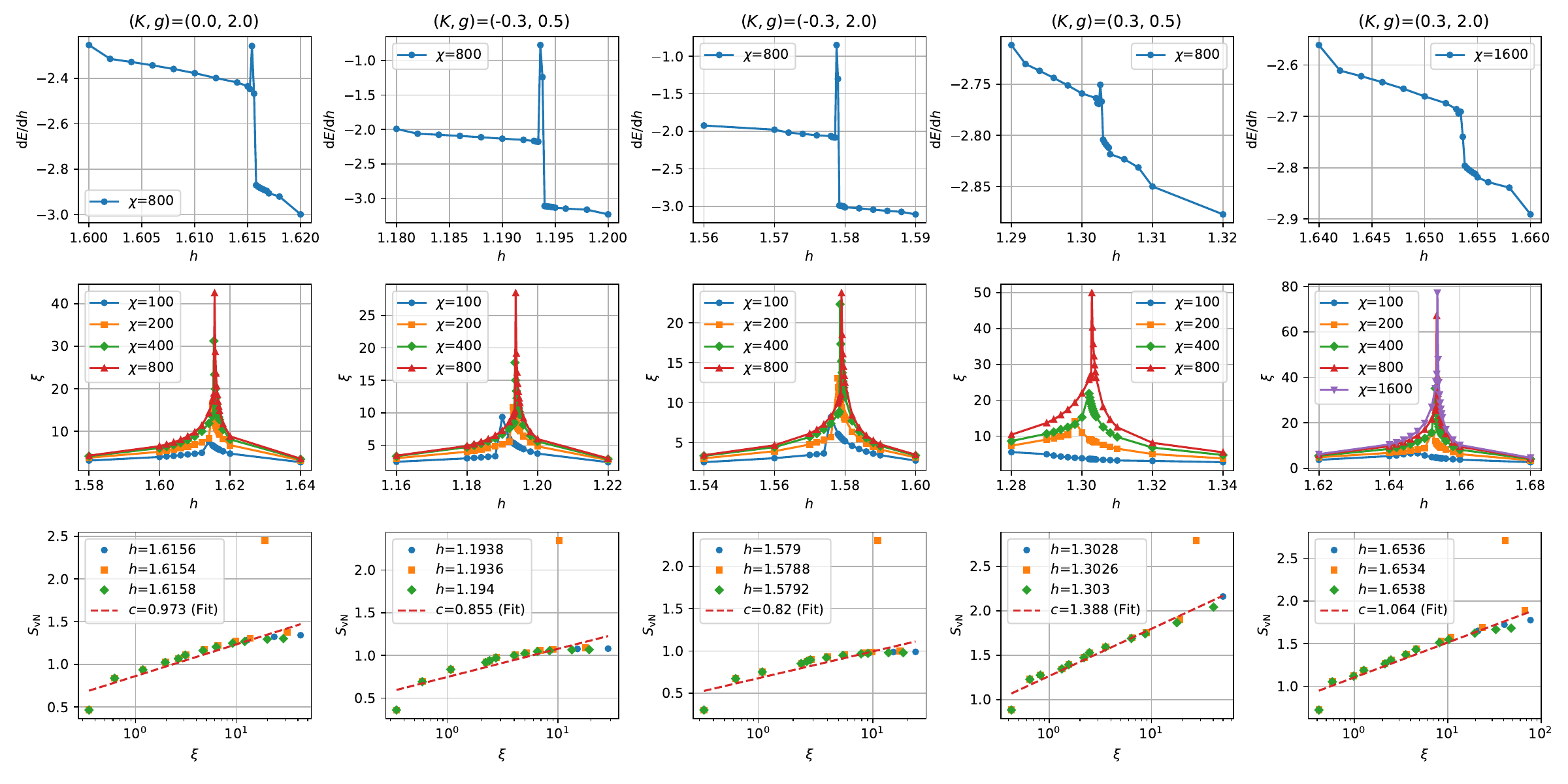}
\caption{
Numerical results for the $N=4$ coupled Ising chains in Eq.~\eqref{eq:NIsingHam} with $(K,g) = (0,2)$, $(-0.3,0.5)$, $(-0.3,2)$, $(0.3,0.5)$, and $(0.3,2)$. 
Top and middle panels show the derivative of the ground-state energy density $dE/dh$ and the correlation length $\xi$ as functions of $h$, respectively. 
Bottom panels show the von Neumann entanglement entropy $S_\textrm{vN}$ plotted against
$\xi$ near the transition points and fitting functions of the form $(c/6) \log \xi + c_0$ (dashed lines).
\label{fig:Coupled4IsingSuppl}
}
\end{figure*}

Finally, in Fig.~\ref{fig:Coupled4IsingSuppl}, we show the derivative of the ground-state energy density $dE/dh$ computed for different choices of the parameter $K$ and $g$ and for the largest available bond dimension $\chi$.
Except for the $K > 0$ cases, it shows a clear discontinuous change across the transition point.
We also compute the correlation length $\xi$ and entanglement entropy $S_\textrm{vN}$.
While the correlation length $\xi$ grows with the bond dimension and shows a sharp peak at the phase transition, the entanglement entropy right at or above the transition shows saturating behavior to a constant value for large $\xi$, expect for $(K,g)=(0.3,0.5)$.
In the latter case, the entanglement entropy right at the transition $h=1.3028$ appears to grow logarithmically for our available values of bond dimension.
These subtle behaviors for $K>0$ might be due to a marginal coupling $\lambda_2$ in the field theory of Eq.~\eqref{eq:NIsingHam}, whose initial value is positive, i.e., $\lambda_2 \sim K > 0$; while $\lambda_2$ eventually flows to $\lambda_2 \to -\infty$ under RG as discussed in Sec.~\ref{sec:RG}, the flow slows down when passing $\lambda_2=0$ and results in a large correlation length, akin to the walking behavior near a complex CFT \cite{Gorbenko18a, Gorbenko18b}.
For the other cases, our numerical results strongly suggest the presence of a first-order transition.

\subsection{SO($N$)-symmetric spin chains}
\label{app:son}

\begin{figure}[tb]
\includegraphics[width=0.48\textwidth]{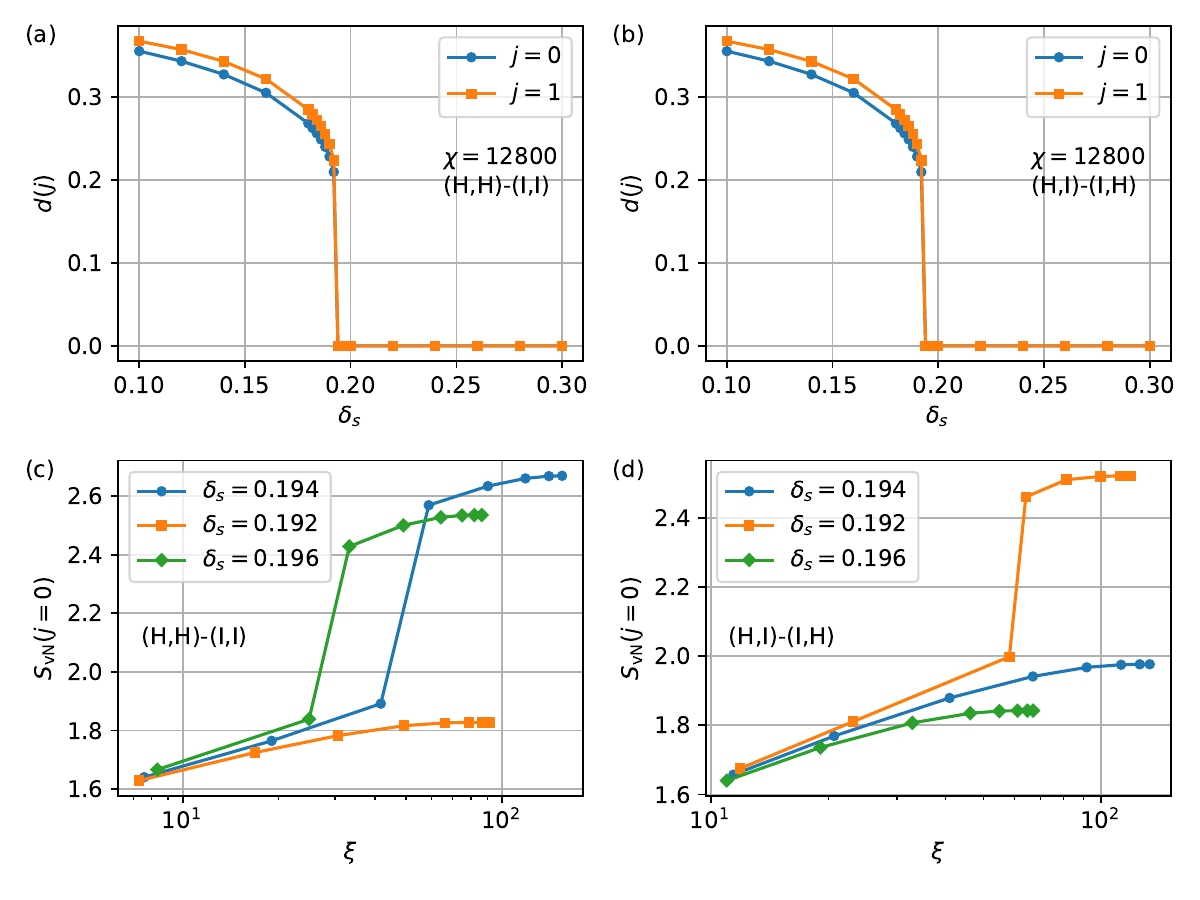}
\caption{
Numerical results for the SO(4)-symmetric ladder in Eq.~\eqref{eq:SO4LadderHam} with $J_1=1$, $J_4=-1$, and $\delta_c=0$.
The dimerization order parameter is computed for the (a) (H,H)-to-(I,I) and (b) (H,I)-to-(I,H) representation of the virtual spaces.
The entanglement entropy is plotted against the correlation length in (c) and (d) for each representation.
\label{fig:SO4LadderSuppl}
}
\end{figure}

We here provide additional numerical results for the SO($4$)-symmetric ladder in Eq.~\eqref{eq:SO4LadderHam}.
In the main text, the representation of the virtual spaces, either (H,H)-to-(I,I) or (H,I)-to-(I,H), is determined by minimizing the ground-state energy density computed from variational MPS with bond dimensions $\chi=12800$. 
For $J_1=1$, $J_4=-1$, and $\delta_c=0$, we have identified the transition point as $\delta_s = 0.194$, across which the representation of the virtual spaces switches.
In fact, as shown in Figs.~\ref{fig:SO4LadderSuppl}(a) and (b), the dimerization order parameter abruptly vanishes across the transition $\delta_s = 0.194$ even if we fix the representation to either (H,H)-to-(I,I) or (H,I)-to-(I,H).
Thus, the abrupt change of the order parameter is not an artifact caused by switching between the two representations but an intrinsic signature of the first-order phase transition. 
In Figs.~\ref{fig:SO4LadderSuppl}(c) and (d), we show the von Neumann entanglement entropy computed around the transition point for each representation.
When we choose a representation that does not minimize the ground-state energy density, the entanglement entropy exhibits a sudden jump as we increase the bond dimension, potentially indicating a level crossing associated with the first-order transition.
In all cases, the entanglement entropy saturates to a constant value for large correlation lengths, which is consistent with a finite excitation gap at the transition.

\clearpage
\twocolumngrid
\bibliography{CoupledIsingTransitions}

\end{document}